# MIMO Multiway Relaying with Clustered Full Data Exchange: Signal Space Alignment and Degrees of Freedom

Xiaojun Yuan, *Member, IEEE*

***Abstract*—We investigate achievable degrees of freedom (DoF) for a multiple-input multiple-output (MIMO) multiway relay channel (mRC) with $L$ clusters and $K$ users per cluster. Each user is equipped with $M$ antennas and the relay with $N$ antennas. We assume a new data exchange model, termed *clustered full data exchange*, i.e., each user in a cluster wants to learn the messages of all the other users in the same cluster. Novel signal alignment techniques are developed to systematically construct the beamforming matrices at the users and the relay for efficient physical-layer network coding. Based on that, we derive an achievable DoF of the MIMO mRC with an arbitrary network configuration of $L$ and $K$, as well as with an arbitrary antenna configuration of $M$ and $N$. We show that our proposed scheme achieves the DoF capacity when $\frac{M}{N} \leq \frac{1}{LK-1}$ and $\frac{M}{N} \geq \frac{(K-1)L+1}{KL}$.**

## I. Introduction

Physical-layer network coding (PNC) has been intensively investigated in the past several years. The earliest model for PNC is the two-way relay channel (TWRC), in which two users exchange information via the help of a single relay [1]. The peer-to-peer based communication protocol for TWRC requires four phases to complete one round of information exchange. PNC reduces the number of the required phases to two by allowing users to transmit or receive signals simultaneously, which implies potentially doubled network throughput.

Since its advent, abundant progresses on the PNC design for TWRC have been reported in the literature; see [1]-[7] and the references therein. Particularly, it was shown in [4] that, with nested lattice coding, the capacity of the TWRC can be achieved within $\frac{1}{2}$ bit. Later, the authors in [5]-[7] considered multiple-input multiple-output (MIMO) TWRCs, in which every node is equipped with multiple antennas. It was revealed that the asymptotic capacity of the MIMO TWRC in the high signal-to-noise (SNR) region can be achieved within a finite gap per spatial dimension for arbitrary antenna configurations.

Tremendous success of PNC over TWRC intrigues intensive research on PNC for more general relay networks. In this regard, a natural generalization of TWRC is called a *multiway relay channel* (mRC), in which users are grouped into clusters and each user in a cluster wants to communicate with other users in the same cluster [8]. This setup generally models a variety of communication scenarios. For example, in a social network, groups of users want to share files via a relay station. Each user in a group only has a distinct portion of a common file desired by all the other users in the same group; many such groups need to be served simultaneously by the relay station. This setup is also relevant to *ad hoc* wireless networks in which nodes that are distributed geographically want to communicate with a central controller to share available local information.

This manuscript is submitted to IEEE Transactions on Wireless Communications. X. Yuan is with the Institute of Network Coding, The Chinese University of Hong Kong, email: xjyuan@inc.cuhk.edu.hk. This work was supported by grants from the University Grants Committee of the Hong Kong Special Administrative Region, China (Project No. 418712 and AoE/E-02/08).

Two special cases of the mRC have been studied in the literature [8]-[12]: in the *full data exchange* model, each user wants to learn the messages from all the other users in the network; in the *pairwise exchange* model, the network consists of multiple pairs, and the two users in each pair want to exchange information with each other. Various relaying protocols and design criteria have been investigated for mRCs operated under these two data exchange models. In particular, it was shown in [8] that the sum-rate capacity of the network can be achieved within a finite-bit gap for any number of clusters.

The initial work on mRC was limited to a single-antenna setup, i.e., each node in the network is equipped with a single antenna. Recently, much attention has been attracted to the MIMO mRC, in which each node in the network is equipped with multiple antennas to allow spatial multiplexing [13]-[20]. For example, the MIMO technique has been introduced into the Y channel in [15] (a special mRC with one cluster and three users per cluster), and also into the multipair TWRC in [14] (a special mRC with multiple clusters and two users in each cluster). An important research avenue on MIMO mRCs is to analyze the degrees of freedom (DoF) of the network, which characterizes the high signal-to-noise ratio (SNR) performance [15]-[20]. It is known that interference alignment, in which interference signals are aligned to occupy a minimal number of temporal/spectral/spatial dimensions, is the key technique to achieve the DoF capacity of various wireless multi-terminal networks [21][22]. As for relay networks, a similar notion, termed *signal space alignment*, was proposed by the authors in [15]. It was shown that, by aligning the signal streams of the users with information exchange to a common direction, the spatial dimensions at the relay can be efficiently utilized, so as to achieve the DoF capacity of the relay network.

This paper considers a general MIMO mRC with $L$ clusters and $K$ users per cluster. Note that a similar channel setup has been previously studied in [17], where the users in a cluster exchange private data in a pairwise manner. In contrast, this paper is focused on a new data exchange model, termed *clustered full data exchange*, in which each user in a cluster wants to learn all the messages from other users in the same cluster. This data exchange model arises in many practical scenarios, such as teleconferencing and data sharing in a social network. The difficulty in the design of efficient communication mechanisms for clustered MIMO mRCs is largely attributed to the following feature of such networks: within each cluster, signal space alignment is necessary to exploit the potential advantage of PNC; meanwhile, for each cluster, the signals from the other clusters are interference, implying that temporal and spatial interference alignment is necessary to minimize the signal dimensions occupied by the interference. Therefore, an efficient protocol over the clustered MIMO mRC involves both network coding and interference alignment, which poses a serious challenge in the system design.

In this paper, we investigate the achievable DoF of the MIMO mRC with clustered full data exchange. We assume a symmetric antenna setup, in which each user is equipped with $M$ antennas and the relay with $N$ antennas. The main contribution of this paper is to develop a novel systematic signal-alignment technique for efficient PNC design over the considered MIMO mRC. Specifically, we call a bunch of $LK$ signal spatial streams as a *unit* if it consists of one spatial signal stream from every user. The signal streams in a unit are aligned to form a certain spatial structure (referred to as a *pattern*) that allows signal separation at the user ends. The number of spatial dimensions occupied by a pattern characterizes the efficiency of this pattern in utilizing the signal space of the relay. Intuitively, a higher ratio of $\frac{M}{N}$ allows higher freedom at the user ends to align the signals, and therefore, a more efficient pattern can be constructed. The signal alignment problem is then to pack as many units (with the most efficient patterns) as possible to occupy the overall signal space of the relay. In this way, the DoF analysis reduces to counting the maximum number of units that can be packed. Based on this technique, an achievable DoF can be derived for the considered MIMO mRC with an arbitrary network configuration of $L$ and $K$, as well as with an arbitrary antenna configuration of $M$ and $N$. By comparing the derived achievable DoF with the cut-set outer bound, we show that the proposed scheme achieves the DoF capacity when $\frac{M}{N} \leq \frac{1}{LK-1}$ and $\frac{M}{N} \geq \frac{(K-1)L+1}{KL}$. We also show that the derived DoF is always piecewise linear and is bounded by either $M$ or $N$, implying that either the users or the relay has redundant antennas. This is similar to the DoF results of the MIMO interference channel obtained in [22].

The remainder of this paper is organized as follows. Section II describes the system model. The achievable DoF of the considered MIMO mRC with $L = 2$ clusters and $K = 3$ users per cluster is derived in Section III. In Section IV, we generalize the results in Section III to an arbitrary configuration of $L$ and $K$. Finally, we close the paper in Section V with some concluding remarks highlighting our main results.

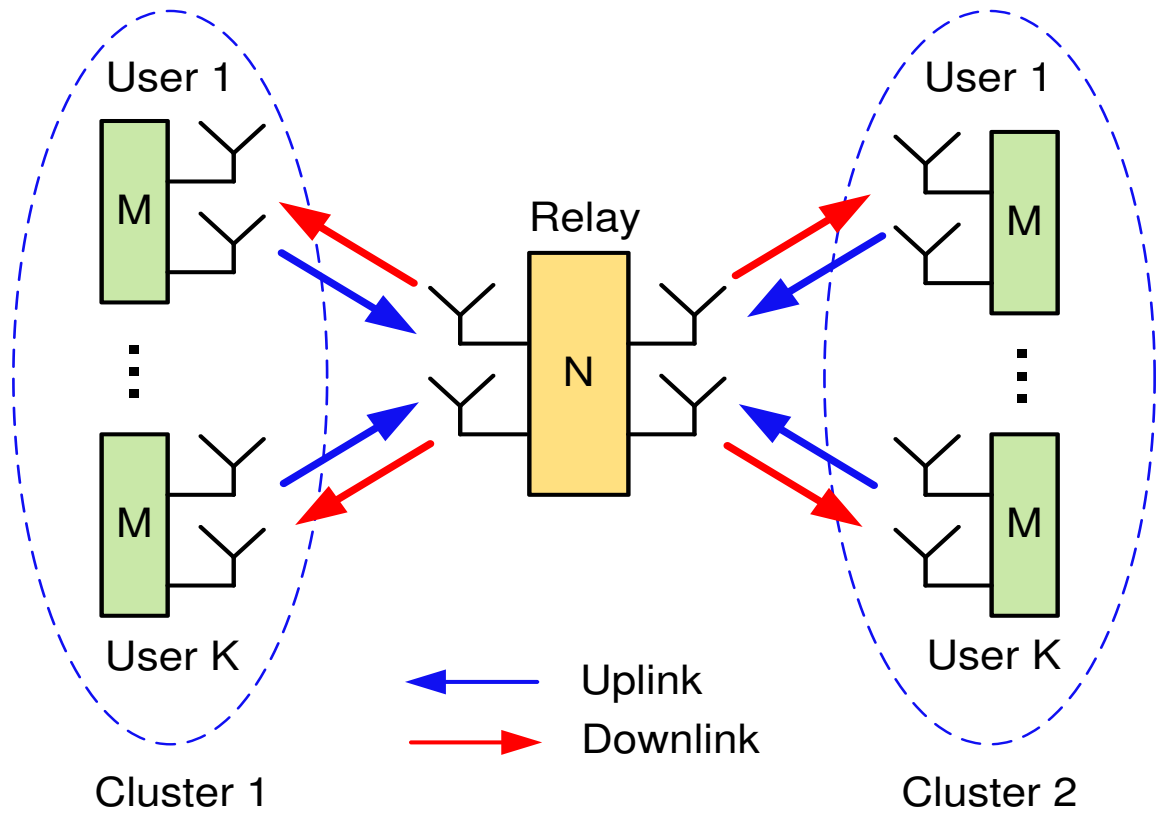

Fig. 1. The MIMO mRC with $L = 2$ clusters and $K$ users in each cluster.

## II. Preliminaries

### A. Notation

The following notation is used throughout this paper. Scalars are denoted by lowercase regular letters, vectors by lowercase bold letters, and matrices by uppercase bold letters. For any matrix $\mathbf{A}$, $\mathbf{A}^T$ and $\mathbf{A}^\dagger$ denote the transpose and the Hermitian transpose, respectively; $\text{span}(\mathbf{A})$ denotes the column space; $\text{null}(\mathbf{A})$ denotes the (right) nullspace; $\text{tr}(\mathbf{A})$ denotes the trace of a square matrix $\mathbf{A}$. $\dim(\mathcal{S})$ denotes the dimension of a space $\mathcal{S}$; $\mathcal{S} \cap \mathcal{U}$ and $\mathcal{S} \oplus \mathcal{U}$ denote the intersection and the direct sum of two spaces $\mathcal{S}$ and $\mathcal{U}$, respectively; $\mathbb{R}^{n \times m}$ and $\mathbb{C}^{n \times m}$ denote the $n$-by-$m$ dimensional real space and complex space, respectively; $(\cdot)^+$ denotes $\max\{\cdot, 0\}$; $\mathcal{CN}(\mu, \sigma^2)$ denotes the circularly symmetric complex Gaussian distribution with mean $\mu$ and variance $\sigma^2$.

### B. System Model

In this paper, we consider a discrete memoryless symmetric MIMO mRC in which multiple users, partitioned into clusters, are simultaneously served by a single relay, and each user in a cluster wants to multicast its message to all other users in the same cluster. We assume that the users cannot overhear each other's transmissions, and that full-duplex communications are employed, i.e., all the users and the relay can transmit and receive signal simultaneously. Let $L$ be the number of clusters and $K$ be the number of users in each cluster. The MIMO mRC with $L = 2$ is illustrated in Fig. 1. Users in cluster $j$, $j \in \mathcal{I}_L \triangleq \{1, ..., L\}$, are denoted by $\mathcal{T}_{j1}, ..., \mathcal{T}_{jK}$. Each user is equipped with $M$ antennas, and the relay with $N$ antennas. For any $k \in \mathcal{I}_K \triangleq \{1, ..., K\}$, the channel matrix from user $k$ of cluster $j$ to the relay is denoted by $\mathbf{H}_{jk} \in \mathbb{C}^{N \times M}$, and the channel matrix from the relay to user $k$ of cluster $j$ is denoted by $\mathbf{G}_{jk}^T \in \mathbb{C}^{M \times N}$. We assume that the elements of $\mathbf{H}_{jk}$ and $\mathbf{G}_{jk}$, $\forall j, k$, are independently drawn from a continuous distribution. Thus the channel matrices are of full column or row rank, whichever is smaller, with probability one. We also assume that the channel state information (CSI) is perfectly known at all nodes.

The Gaussian MIMO mRC is modeled as

$$\mathbf{Y}_R = \sum_{j=1}^{L}\sum_{k=1}^{K}\mathbf{H}_{jk}\mathbf{X}_{jk} + \mathbf{Z}_R \quad (1a)$$

$$\mathbf{Y}_{jk} = \mathbf{G}_{jk}^T\mathbf{X}_R + \mathbf{Z}_{jk}, \quad j \in \mathcal{I}_L, k \in \mathcal{I}_K, \quad (1b)$$

where $\mathbf{X}_{jk} \in \mathbb{C}^{M\times T}$ and $\mathbf{Y}_{jk} \in \mathbb{C}^{M\times T}$ are the input and the output at user $\mathcal{T}_{jk}$, respectively; $\mathbf{X}_R \in \mathbb{C}^{N\times T}$ and $\mathbf{Y}_R \in \mathbb{C}^{N\times T}$ are the input and the output at the relay, respectively; $T$ denotes the number of channel uses in a transmission frame; $\mathbf{Z}_R \in \mathbb{C}^{N\times T}$ and $\mathbf{Z}_{jk} \in \mathbb{C}^{M\times T}$, respectively, are the additive white Gaussian noise (AWGN) matrices at the relay and at user $\mathcal{T}_{jk}$, with the elements independently drawn from $\mathcal{CN}(0,\sigma^2)$. For the considered symmetric network, the power constraints of the transmitted signals at user $\mathcal{T}_{jk}$ and at the relay are respectively given by

$$\tfrac{1}{T}\mathrm{tr}(\mathbf{X}_{jk}\mathbf{X}_{jk}^\dagger) \le P, \quad j \in \mathcal{I}_L, k \in \mathcal{I}_K \quad (2a)$$

$$\tfrac{1}{T}\mathrm{tr}(\mathbf{X}_R\mathbf{X}_R^\dagger) \le P \quad (2b)$$

where $P$ is the maximum transmission power allowed at the relay and at the users.

### C. Linear Processing at Users and Relay

We now present detailed operations at the users and the relay. Throughout this paper, we assume *clustered full data exchange*. That is, each user $\mathcal{T}_{jk}$, $j \in \mathcal{I}_L$, $k \in \mathcal{I}_K$, sends a common message $W_{jk}$ to all the other users in cluster $j$, and wants to learn the messages $W_{jk'}$, $k' \in \mathcal{I}_K\backslash\{k\}$, from all the other users in cluster $j$.

We first consider the uplink phase. Let $\mathbf{X}'_{jk} \in \mathbb{C}^{m\times T}$ be the codeword matrix of user $\mathcal{T}_{jk'}$ one-to-one mapped to $W_{jk}$, where $m$ represents the number of independent spatial streams with $m \le M$. Denote by $\mathbf{U}_{jk} \in \mathbb{C}^{M\times m}$ the corresponding precoding matrix. Then, the channel input of user $\mathcal{T}_{jk'}$ is

$$\mathbf{X}_{jk} = \mathbf{U}_{jk}\mathbf{X}'_{jk}, \quad j \in \mathcal{I}_L, k \in \mathcal{I}_K. \quad (3)$$

The relay operation is described as follows. For convenience, we introduce the following notation:

$$\mathbf{M}_j = [\mathbf{H}_{j1}\mathbf{U}_{j1},\cdots,\mathbf{H}_{jK}\mathbf{U}_{jK}], \quad j \in \mathcal{I}_L \quad (4a)$$

$$\mathbf{M} = [\mathbf{M}_1,\cdots,\mathbf{M}_L] \quad (4b)$$

$$\mathbf{M}_{\bar{j}} = [\mathbf{M}_1,\cdots,\mathbf{M}_{j-1},\mathbf{M}_{j+1},\cdots,\mathbf{M}_L], \; j \in \mathcal{I}_L. \quad (4c)$$

For each cluster $j$, the relay extracts the signal component of $\mathbf{Y}_R$ orthogonal to the signals of the other clusters. To this end, let $\mathbf{P}_j \in \mathbb{C}^{N\times N}$ be the projection matrix that projects a vector into the subspace $\mathrm{null}(\mathbf{M}_{\bar{j}})$. Then, for each cluster $j$, the relay obtains

$$\mathbf{P}_j\mathbf{Y}_R = \mathbf{P}_j\mathbf{H}_j\mathbf{X}'_j + \mathbf{P}_j\mathbf{Z}_R \quad (5a)$$

where $\mathbf{X}'_j = [\mathbf{X}'_{j1},\cdots,\mathbf{X}'_{jK}]^T$. In the above, the signals from cluster $j' \ne j$ disappear due to the projection.

We now consider the downlink phase. Similarly to (4), we denote $\mathbf{N}_j = [\mathbf{G}_{j1}\mathbf{V}_{j1},\cdots,\mathbf{G}_{jK}\mathbf{V}_{jK}]$, and $\mathbf{N}_{\bar{j}} = [\mathbf{N}_1,\cdots,\mathbf{N}_{j-1},\mathbf{N}_{j+1},\cdots,\mathbf{N}_L]$, where $\mathbf{V}_{jk} \in \mathbb{C}^{M\times m}$ is the receive processing matrix of user $\mathcal{T}_{jk'}$. Let $\mathbf{W}_j \in \mathbb{C}^{N\times N}$ be the projection matrix that projects a vector into the subspace $\mathrm{null}(\mathbf{N}_{\bar{j}})$. The relay sends out the signal $\mathbf{W}_j^T\mathbf{P}_j\mathbf{Y}_R$ for each cluster $j$, i.e., the relay's transmit signal is given by

$$\mathbf{X}_R = \sum_{j=1}^{L}\mathbf{W}_j^T\mathbf{P}_j\mathbf{Y}_R. \quad (6)$$

The relay-to-user signal model is then given by

$$\mathbf{V}_{jk}^T\mathbf{Y}_{jk} = \mathbf{V}_{jk}^T\mathbf{G}_{jk}^T\sum_{j'=1}^{L}\mathbf{W}_{j'}^T\mathbf{P}_{j'}\mathbf{Y}_R + \mathbf{V}_{jk}^T\mathbf{Z}_{jk} \quad (7a)$$

$$= \mathbf{V}_{jk}^T\mathbf{G}_{jk}^T\mathbf{W}_j^T\mathbf{P}_j\mathbf{Y}_R + \mathbf{V}_{jk}^T\mathbf{Z}_{jk} \quad (7b)$$

$$= \mathbf{V}_{jk}^T\mathbf{G}_{jk}^T\mathbf{W}_j^T\mathbf{P}_j\left(\sum_{k'=1}^{K}\mathbf{H}_{jk'}\mathbf{U}_{jk'}\mathbf{X}'_{jk'} + \mathbf{Z}_R\right) + \mathbf{V}_{jk}^T\mathbf{Z}_{jk}. \quad (7c)$$

Upon receiving $\mathbf{V}_{jk}^T\mathbf{Y}_{jk}$, each user $\mathcal{T}_{jk}$ computes the message estimate of $W_{jk'}$, denoted by $\hat{W}_{jk}^{k'}$, for $k' = 1,\cdots,k-1,k+1,\cdots,K$.

We note from (7) that the equivalent channel from user $\mathcal{T}_{jk'}$ to user $\mathcal{T}_{jk}$ is given by $\mathbf{V}_{jk}^T\mathbf{G}_{jk}^T\mathbf{W}_j^T\mathbf{P}_j\mathbf{H}_{jk'}\mathbf{U}_{jk'}$. This equivalent channel can be splitted into two symmetric components, namely, the uplink one $\mathbf{P}_j\mathbf{H}_{jk'}\mathbf{U}_{jk'}$ and the downlink one $\mathbf{V}_{jk}^T\mathbf{G}_{jk}^T\mathbf{W}_j^T$. This symmetry implies that any beamforming design in the uplink phase carries directly over to the downlink phase. Therefore, we mostly focus on the uplink beamforming design in what follows.

### D. Degrees of Freedom

This paper focuses on analyzing the DoF of the considered MIMO mRC. Roughly speaking, the DoF of a network is the number of independent signal streams that can be supported by the network. To make this notion rigorous, we introduce the following definitions.

Let $R_{jk}$ be the information rate carried in $W_{jk}$. We say that user $\mathcal{T}_{jk'}$ achieves a sum rate of $C_{jk} = \sum_{k=1,k\ne k'}^{K} R_{jk}$, if $\Pr\{\hat{W}_{jk}^{k'} \ne W_{jk}\}$ tends to zero for $k \in \mathcal{I}_K\backslash\{k'\}$ as $T \to \infty$. Denote $SNR = P/\sigma^2$. The achievable sum rate $C_{jk}$ is in general a function of $SNR$, denoted as $C_{jk}(SNR)$, $j \in \mathcal{I}_L$, $k \in \mathcal{I}_K$. We define the corresponding total achievable DoF as

$$d_{\mathrm{sum}} \triangleq \lim_{SNR\to\infty}\frac{\sum_{j=1}^{L}\sum_{k=1}^{K}C_{jk}(SNR)}{\log SNR} \quad (8)$$

where $C_{jk}(SNR)$ is in bit, and "log" denotes logarithm with base 2. Also, we define the corresponding achievable DoF per user as

$$d_{\mathrm{user}} \triangleq \frac{1}{KL}d_{\mathrm{sum}} \quad (9)$$

and the achievable DoF per relay dimension as

$$d_{\mathrm{relay}} \triangleq \frac{1}{N}d_{\mathrm{sum}}. \quad (10)$$

Later, we will see that $d_{\mathrm{relay}}$ measures the efficiency of utilizing the signal space at the relay.

A cut-set outer bound on the DoF of the considered MIMO mRC is given as

$$d_{\mathrm{sum}} \le \min(KLM, KN), \quad (11a)$$

or equivalently

$$d_{\text{user}} \leq \min\left(M, \frac{N}{L}\right). \tag{11b}$$

The above outer bound can be intuitively explained as follows. On one hand, each user has $M$ antennas and can decode at most $M$ independent data streams. Thus, $d_{\text{user}}$ is upper-bounded by $M$. On the other hand, the relay has $N$ antennas and simultaneously serves $L$ clusters. Thus, the relay can deliver at most $\frac{N}{L}$ independent data streams to the users in each cluster. This outer bound will be used as a benchmark in the following analysis. Also, the DoF capacity is achieved when an achievable DoF meets the outer bound.

## III. The Case of $L = 2$ and $K = 3$

In this section, we consider a MIMO mRC with $L = 2$ clusters and $K = 3$ users in each cluster. We will generalize our results to arbitrary values of $L$ and $K$ in the next section.

### A. Preliminary Discussions

We start with some intuitions on signal space alignment for a MIMO mRC. We refer to the $l$-th row of the codeword matrix $\mathbf{X}'_{jk}$, denoted by $\mathbf{x}'^{(l)}_{jk}$, as the $l$-th spatial stream of user $\mathcal{T}_{jk}$. From (1) and (3), we see that $\mathbf{x}'^{(l)}_{jk}$ impinges upon the relay in the direction of $\mathbf{h}^{(l)}_{jk} \triangleq \mathbf{H}_{jk}\mathbf{u}^{(l)}_{jk}$, where $\mathbf{u}^{(l)}_{jk}$ is the $l$-th column of $\mathbf{U}_{jk}$. Similarly, let $\mathbf{v}^{(l)}_{jk}$ be the $l$-th column of $\mathbf{V}_{jk}$. Then, the relay transmits the $l$-th spatial stream to user $\mathcal{T}_{jk}$ in the direction of $\mathbf{g}^{(l)}_{jk} \triangleq \mathbf{G}_{jk}\mathbf{v}^{(l)}_{jk}$. We refer to the $l$-th spatial streams of all users, i.e., $\{\mathbf{x}'^{(l)}_{jk}|\forall j,k\}$, as a unit. Clearly, a unit in general contains $KL$ spatial streams, one from each user. Further, we refer to the spatial structure formed by $\{\mathbf{h}^{(l)}_{jk}|\forall j,k\}$ and $\{\mathbf{g}^{(l)}_{jk}|\forall j,k\}$ as a pattern.

We design a pattern in such a way that each user achieves one DoF. Suppose that each user $\mathcal{T}_{jk}$ transmits only one spatial stream. From (7), we see that each user $\mathcal{T}_{jk}$ receives one linear equation with the equivalent channel coefficient for the link between $\mathcal{T}_{jk}$ and $\mathcal{T}_{jk'}$ given by $\mathbf{g}^T_{jk}\mathbf{W}^T_j\mathbf{P}_j\mathbf{h}_{jk'}$.[1] Note that $\mathbf{g}_{jk}$, $\mathbf{W}_j$, $\mathbf{P}_j$, $\mathbf{h}_{jk'}$ are statistically independent of each other. Thus, $\mathbf{g}^T_{jk}\mathbf{W}^T_j\mathbf{P}_j\mathbf{h}_{jk'}$ is nonzero with probability one, provided that both $\text{null}(\mathbf{M}_{\bar{j}})$ and $\text{null}(\mathbf{N}_{\bar{j}})$ are of at least dimension one (or equivalently, $\text{rank}(\mathbf{P}_j) \geq 1$ and $\text{rank}(\mathbf{W}_j) \geq 1$). In this way, each user $\mathcal{T}_{jk}$ obtains one linear combination of the signals from the other $K-1$ users in cluster $j$ in one channel use (after self-interference cancellation). Combining $K-1$ channel uses, each user in cluster $j$ has $K-1$ independent combinations and is able to decode $K-1$ messages from the other users in the same cluster, achieving a DoF of $d_{\text{user}} = 1$ per channel use.

The following five patterns satisfy the above design criteria for the case of $L = 2$ and $K = 3$.

1) **Pattern 1.1:** $\{\mathbf{h}_{jk}|j \in \{1,2\}, k \in \mathcal{I}_K\}$ span a subspace of dimension 6 (dim-6), and so do $\{\mathbf{g}_{jk}|j \in \{1,2\}, k \in \mathcal{I}_K\}$.

[1]Here the unit index in $\mathbf{h}_{jk'}$ and $\mathbf{g}_{jk'}$ is omitted, as there is only one unit considered in the design.

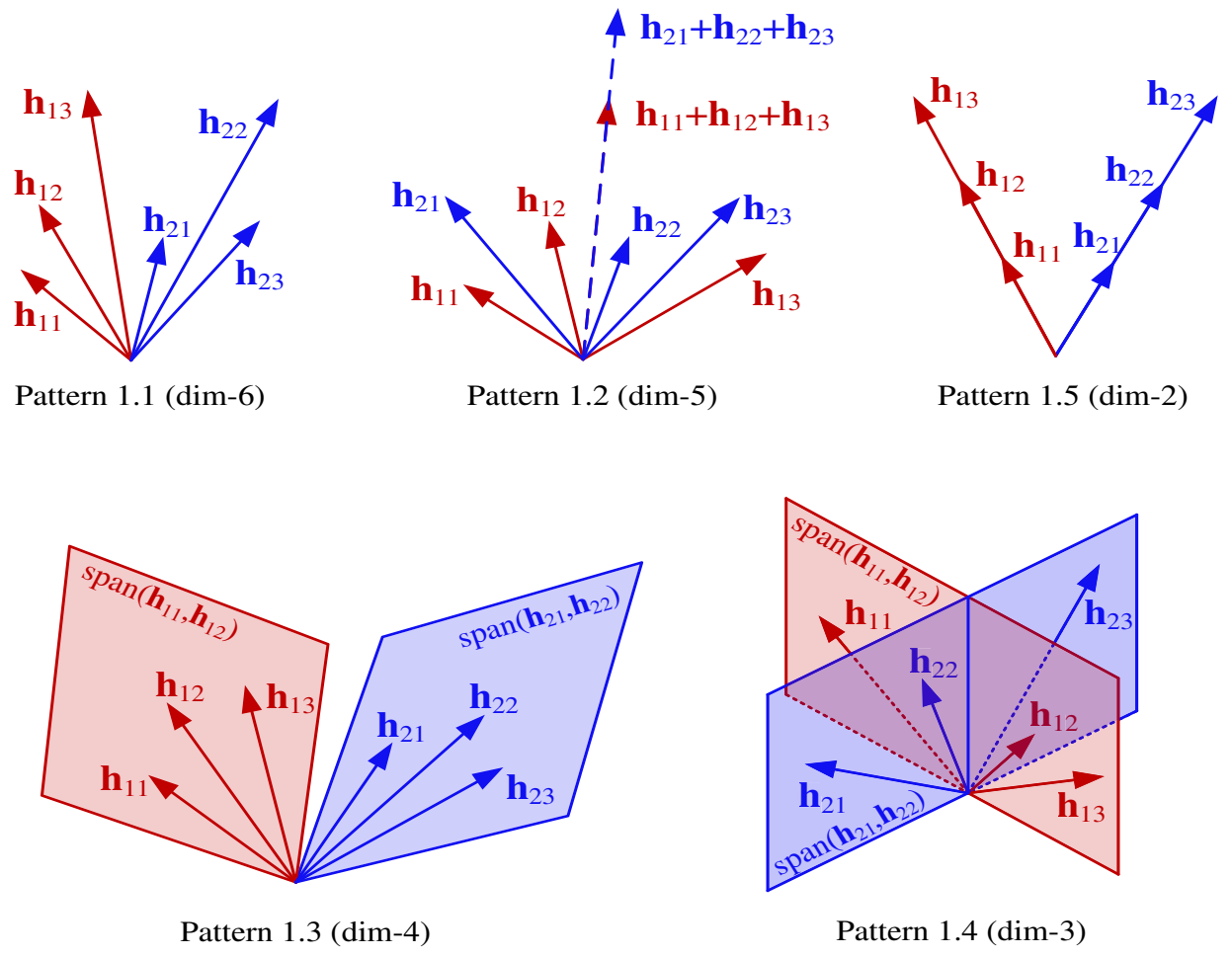

Fig. 2. An illustration of Patterns 1.1 to 1.5.

2) **Pattern 1.2:** $\{\mathbf{h}_{jk}|j \in \{1,2\}, k \in \mathcal{I}_K\}$ span a subspace of dim-5, and so do $\{\mathbf{g}_{jk}|j \in \{1,2\}, k \in \mathcal{I}_K\}$; for any cluster $j$, $\{\mathbf{h}_{jk}|k \in \mathcal{I}_K\}$ span a dim-3 subspace, and so do $\{\mathbf{g}_{jk}|k \in \mathcal{I}_K\}$.
3) **Pattern 1.3:** $\{\mathbf{h}_{jk}|j \in \{1,2\}, k \in \mathcal{I}_K\}$ span a subspace of dim-4, and so do $\{\mathbf{g}_{jk}|j \in \{1,2\}, k \in \mathcal{I}_K\}$; for any cluster $j$, $\{\mathbf{h}_{jk}|k \in \mathcal{I}_K\}$ span a dim-2 subspace, and so do $\{\mathbf{g}_{jk}|k \in \mathcal{I}_K\}$.
4) **Pattern 1.4:** $\{\mathbf{h}_{jk}|j \in \{1,2\}, k \in \mathcal{I}_K\}$ span a subspace of dim-3, and so do $\{\mathbf{g}_{jk}|j \in \{1,2\}, k \in \mathcal{I}_K\}$; for any cluster $j$, $\{\mathbf{h}_{jk}|k \in \mathcal{I}_K\}$ span a dim-2 subspace, and so do $\{\mathbf{g}_{jk}|k \in \mathcal{I}_K\}$.
5) **Pattern 1.5:** $\{\mathbf{h}_{jk}|j \in \{1,2\}, k \in \mathcal{I}_K\}$ span a subspace of dim-2, and so do $\{\mathbf{g}_{jk}|j \in \{1,2\}, k \in \mathcal{I}_K\}$; for any cluster $j$, $\{\mathbf{h}_{jk}|k \in \mathcal{I}_K\}$ span a dim-1 subspace, and so do $\{\mathbf{g}_{jk}|k \in \mathcal{I}_K\}$.

An illustration of the above five patterns is given in Fig. 2, where only the uplink channel vectors are demonstrated by noting the uplink/downlin symmetry. It can been seen that, for these patterns, the corresponding projection matrices $\{\mathbf{W}_j\}$ and $\{\mathbf{P}_j\}$ are of at least rank one. As an example, for Pattern 1.2, the overall signal space occupied by $\{\mathbf{h}_{jk}|j \in \{1,2\}, k \in \mathcal{I}_K\}$ is of dim-5. Each cluster $j$ occupies a signal subspace of dim-3. Thus, $\text{null}(\mathbf{M}_{\bar{j}})$ is at least of dim-2, i.e., $\mathbf{P}_j$ is at least of rank 2. The reasoning for the other patterns are similar. Then, from (7) and the discussions therein, we conclude that each of the above five patterns can achieve the same total DoF of 6. Nevertheless, the dimension of the relay's signal space spanned by these channel vectors differs from pattern to pattern, which leads to a varying DoF per relay dimension (i.e., $d_{\text{relay}}$); see Table I. Clearly, the greater $d_{\text{relay}}$ is, the more efficiently the relay's signal space is utilized. Thus, the priority of a pattern is ranked by $d_{\text{relay}}$, with Pattern 1.5 of the highest priority. Besides, we note that the last column of Table I gives the requirement on the antenna setup to construct these patterns in a MIMO mRC. The details on how to obtain these requirements will be elaborated in Subsection III-C.

TABLE I
PATTERNS FOR THE MIMO MRC WITH $L=2$ AND $K=3$

| Pattern | Dimension | $d_{\text{sum}}$ | $d_{\text{relay}}$ | Requirement |
|---|---|---|---|---|
| 1.1 | 6 | 6 | 1 | N.A. |
| 1.2 | 5 | 6 | $\frac{6}{5}$ | $\frac{M}{N} > \frac{1}{6}$ |
| 1.3 | 4 | 6 | $\frac{3}{2}$ | $\frac{M}{N} > \frac{1}{3}$ |
| 1.4 | 3 | 6 | 2 | $\frac{M}{N} \geq \frac{4}{9}$ |
| 1.5 | 2 | 6 | 3 | $\frac{M}{N} > \frac{2}{3}$ |

## B. Main Result

We now consider the general case that each user transmits multiple spatial streams over a MIMO mRC with $L=2$ and $K=3$. We will design the beamforming matrices $\{\mathbf{U}_{jk}\}$ and $\{\mathbf{V}_{jk}\}$ to align signals in such a way that the equivalent channel vectors $\{\mathbf{H}_{jk}\mathbf{u}_{jk}^{(l)}|\forall j,k\}$ and $\{\mathbf{G}_{jk}\mathbf{v}_{jk}^{(l)}|\forall j,k\}$ form one of the five patterns described in Subsection III-A. Further, as aforementioned, the priorities of these patterns are ranked by $d_{\text{relay}}$. Our target is to determine the most efficient way of constructing units to occupy the relay signal space, with the result given below.

*Lemma 1:* For the $M$-by-$N$ MIMO mRC with $L=2$ and $K=3$ operating in the clustered full data exchange mode, an achievable DoF per user is given by

$$d_{\text{user}} = \begin{cases} \min\left(M, \frac{N}{5}\right), & \frac{M}{N} \leq \frac{1}{3} \\ \min\left(\frac{3M}{5}, \frac{N}{4}\right), & \frac{1}{3} < \frac{M}{N} < \frac{4}{9} \\ \frac{N}{3}, & \frac{4}{9} \leq \frac{M}{N} \leq \frac{2}{3} \\ \min\left(M - \frac{N}{3}, \frac{N}{2}\right), & \frac{M}{N} > \frac{2}{3}. \end{cases} \quad (12)$$

The proof of Lemma 1 can be found in the next subsection. From Lemma 1, we observe the following interesting property.

*Property 1:* The achievable DoF is proportional to the relay's antenna number $N$ for any given antenna ratio $\frac{M}{N}$. That is, $d_{\text{user}} = Nf\left(\frac{M}{N}\right)$, where $f(\cdot)$ is a function of $\frac{M}{N}$. This property generally holds for MIMO mRCs by noting the fact that doubling both $M$ and $N$ always doubles the DoF of the MIMO mRC. This property will be used to prove an important lemma shortly.

The DoF results in Lemma 1 is obtained by aligning signals in the $N$-dimension signal space at the relay. All the $N$ antennas at the relay are utilized in signaling. Physically, it is always allowed to disable a portion of antennas at any node. We next show that, for certain configurations of $M$ and $N$, disabling a portion of relay antennas (and aligning signals in a relay's signal space with a smaller dimension) improves the achievable DoF. To show this, we first present the antenna-disablement lemma below.

*Lemma 2:* For the considered MIMO mRC, assume that a DoF of $d_{\text{user}} = d_0$ is achievable at a certain antenna configuration of $(M=M_0, N=N_0)$. Then, every point of $\left(x = \frac{M}{N_0}, y = d_{\text{user}}\right)$ on the line segment of $y = d_0$ for $x \in [\frac{M_0}{N_0}, \infty)$ is achievable by disabling $M - M_0$ user antennas; also, every point of $\left(x = \frac{M}{N_0}, y = d_{\text{user}}\right)$ on the line segment of $y = \frac{d_0 N_0}{M_0}x$ for $x \in (0, \frac{M_0}{N_0}]$ is achievable by disabling $N_0 - \frac{MN_0}{M_0}$ relay antennas.

*Proof:* The first half of the lemma (i.e., the statement on disabling user antennas) trivially holds. We focus on the second half. By assumption, $d_{\text{user}} = d_0$ is achievable at $(M=M_0, N=N_0)$. Then, from Property 1, $d_{\text{user}} = \frac{d_0}{N_0}N$ is achievable at any $(M,N)$ satisfying $\frac{M}{N} = \frac{M_0}{N_0}$.

Now consider an antenna configuration $(M, N=N_0)$ with $M < M_0$. To prove Lemma 2, it suffices to show that the DoF of $d_{\text{user}} = \frac{d_0 M}{M_0}$ is achievable. As $\frac{M}{N_0} < \frac{M_0}{N_0}$, we can reduce the number of active relay antennas to $N' = \frac{MN_0}{M_0}$ by disabling $N_0 - \frac{MN_0}{M_0}$ relay antennas. Then, $\frac{M}{N'} = \frac{M_0}{N_0}$, implying that a DoF of $d_{\text{user}} = \frac{d_0}{N_0}N' = \frac{d_0 M}{M_0}$ is achievable. This completes the proof of Lemma 2. ■

Combining Lemmas 1 and 2, we obtain the following main result of this section.

*Theorem 3:* For the considered $M$-by-$N$ MIMO mRC with $L=2$ and $K=3$, an improved achievable DoF per user is given by

$$d_{\text{user}} = \begin{cases} \min\left(M, \frac{N}{5}\right), & \frac{M}{N} \leq \frac{4}{15} \\ \min\left(\frac{3M}{4}, \frac{N}{3}\right), & \frac{4}{15} < \frac{M}{N} < \frac{5}{9} \\ \min\left(\frac{3M}{5}, \frac{N}{2}\right), & \frac{M}{N} > \frac{5}{9}. \end{cases} \quad (13)$$

The function of the achievable DoF against $\frac{M}{N}$ is plotted in Fig. 3. It is interesting to see how the achievable DoF given by Lemma 1 is improved by using Lemma 2. As illustrated, the relay antenna disablement lemma increases the achievable DoF in both ranges of $\frac{M}{N} \in (\frac{4}{15}, \frac{4}{9})$ and $\frac{M}{N} \in (\frac{5}{9}, \frac{5}{6})$. Also, we see that the improved achievable DoF curve is piecewise linear and is bounded either by the user antenna number $M$ or by the relay antenna number $N$, which is analogous to the DoF results for the MIMO interference channel [22].

In Fig. 3, we also include the cut-set outer bound for comparison. We see that the achievable bound and the outer bound coincide when $\frac{M}{N} \geq \frac{5}{6}$ and $\frac{M}{N} \leq \frac{1}{5}$, implying that the DoF capacity is achieved in these ranges of $\frac{M}{N}$ values. However, for $\frac{1}{5} < \frac{M}{N} < \frac{5}{6}$, there is still a gap between the achievable DoF and the outer bound. Reducing this gap will be an interesting topic for future research.

## C. Proof of Lemma 1

As aforementioned, we need to jointly design the transmit beamforming vectors $\{\mathbf{u}_{jk}^{(l)}|\forall j, \forall k, \forall l\}$, the receive beamforming vectors $\{\mathbf{v}_{jk}^{(l)}|\forall j, \forall k, \forall l\}$, and the relay's projection matrices, so as to achieve the DoF specified in Lemma 1. We will mostly focus on the uplink beamforming design, as the downlink design is straightforward by noting the uplink/downlink symmetry. Our goal is to design as many units as possible

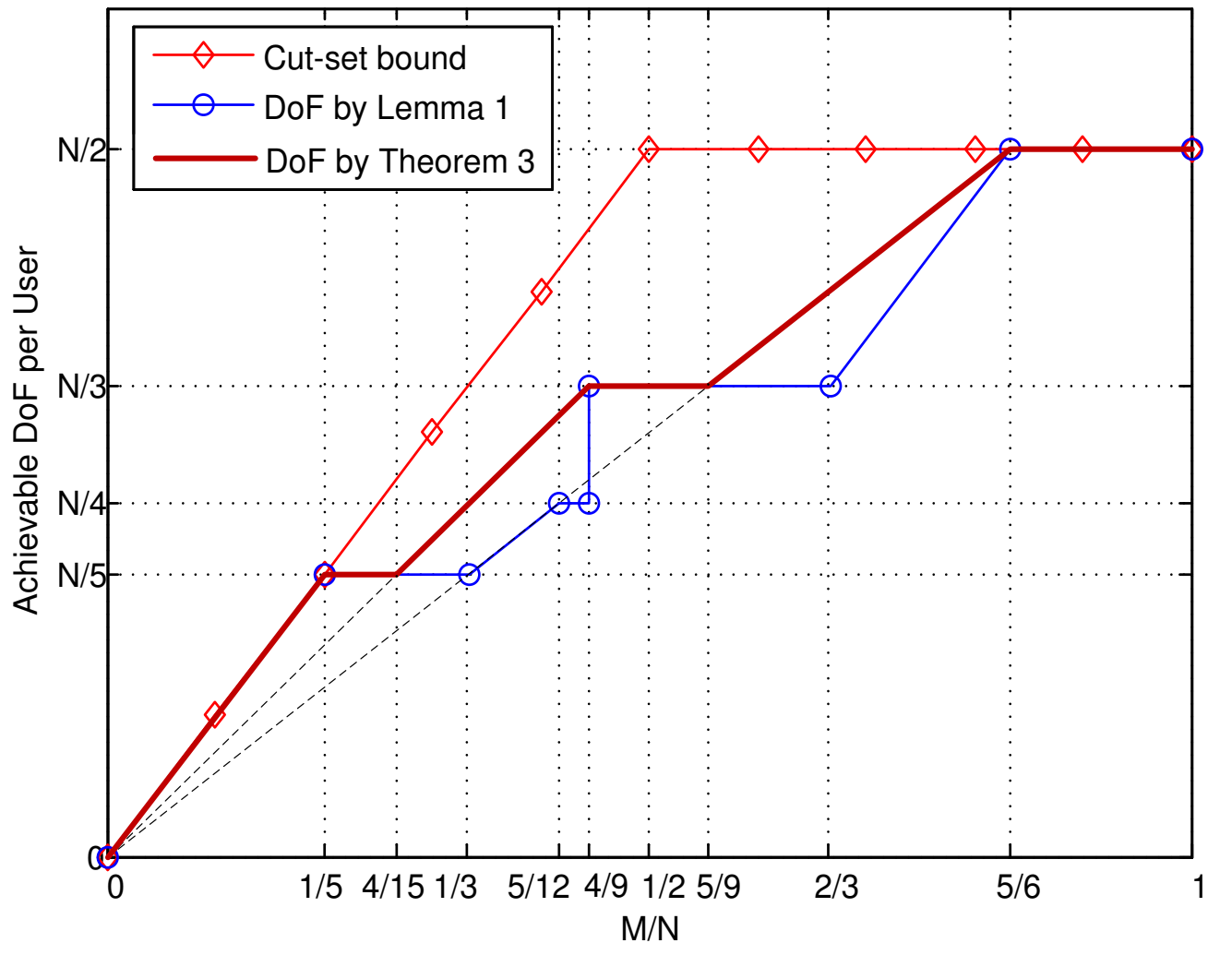

Fig. 3. The achievable DoF per user and the DoF cut-set bound with respect to the antenna ratio $\frac{M}{N}$ for the 2-cluster MIMO mRC operating in the clustered full data exchange model. Each cluster has 3 users.

to pack the relay's signal space, under the constraint that the messages in each unit are decodable at their intended user ends.

*1) Case of $\frac{M}{N} \leq \frac{1}{6}$:* As $N \geq 6M$, both $\text{span}(\{\mathbf{H}_{jk}|\forall j, \forall k\})$ and $\text{span}(\{\mathbf{G}_{jk}|\forall j, \forall k\})$ are of dim-$6M$ with probability one. This implies that the relay has enough signal space to allow each user to transmit $M$ spatial streams, which amounts to $M$ units with Pattern 1.1. Thus, an achievable DoF per user is $M$.

*2) Case of $\frac{1}{6} < \frac{M}{N} \leq \frac{1}{3}$:* In this case, $\frac{M}{N}$ is sufficiently large to construct Pattern 1.2. Denote $\mathbf{H} = [\mathbf{H}_{11}, \mathbf{H}_{12}, \mathbf{H}_{13}, \mathbf{H}_{21}, \mathbf{H}_{22}, \mathbf{H}_{23}] \in \mathbb{C}^{N \times 6M}$. From channel randomness, $\text{null}(\mathbf{H})$ is of dimension $6M - N$ with high probability. Thus, with probability one, there exists a full column-rank matrix $\mathbf{U} \in \mathbb{C}^{6M \times (6M-N)}$ satisfying

$$\mathbf{H}\mathbf{U} = \mathbf{0}, \tag{14}$$

or equivalently

$$\sum_{j=1}^{L} \sum_{k=1}^{K} \mathbf{H}_{jk}\mathbf{U}_{jk} = \mathbf{0}, \tag{15}$$

where $\mathbf{U}$ is partitioned as $\mathbf{U} = \left[\mathbf{U}_{11}^T, \mathbf{U}_{12}^T, \mathbf{U}_{13}^T, \mathbf{U}_{21}^T, \mathbf{U}_{22}^T, \mathbf{U}_{23}^T\right]^T$ with $\mathbf{U}_{jk} \in \mathbb{C}^{M \times (6M-N)}$, $\forall j, k$. From (15), the rank of $\mathbf{M} \in \mathbb{C}^{N \times 6(6M-N)}$ (defined in (4b))is at most $\min\{N, 5(6M - N)\}$. From the randomness of $\mathbf{H}$, $\mathbf{U}_{jk}$ is of full rank for sure. Then, using Lemma 10 in Appendix A, we further see that $\mathbf{M}$ is of rank $\min\{N, 5(6M - N)\}$ with probability one.

First consider $5(6M - N) < N$, or equivalently, $\frac{M}{N} < \frac{1}{5}$. Denote by $\mathbf{u}_{jk}^{(l)}$ the $l$-th column of $\mathbf{U}_{jk}$. Then

$$\sum_{j=1}^{L} \sum_{k=1}^{K} \mathbf{H}_{jk}\mathbf{u}_{jk}^{(l)} = \mathbf{0}, \text{for } l = 1, \cdots, 6M - N. \tag{16}$$

Thus, $\text{span}(\{\mathbf{H}_{jk}\mathbf{u}_{jk}^{(l)}|\forall j, k\})$ is of at most dim-5, for $\forall l$. As $\text{rank}(\mathbf{M}) = 5(6M - N)$, $\text{span}(\{\mathbf{H}_{jk}\mathbf{u}_{jk}^{(l)}|\forall j, k\})$ is of dim-5 with probability one, for $\forall l$. Therefore, $\{\mathbf{H}_{jk}\mathbf{u}_{jk}^{(l)}|\forall j, k\}$ forms a unit with Pattern 1.2, for $l = 1, \cdots, 6M - N$. These $6M - N$ units occupy a dim-$5(6M - N)$ subspace with probability one. The remaining $N - 5(6M - N) = 6(N - 5M)$ dimensions are used for constructing $N - 5M$ units with Pattern 1.1.[2] Specifically, each user has $M - (6M - N) = N - 5M$ unused dimensions. In other words, each user can transmit $N - 5M$ extra data streams at most. Hence the total number of the spatial streams transmitted by all the users is $6(N - 5M)$, which is exactly the number of the unused dimensions at the relay. Thus, we can construct $N - 5M$ units with Pattern 1.1. Then, each user transmits $M$ spatial streams in total. Therefore, the achievable DoF per user is $M$.

We now consider the remaining case of $\frac{M}{N} \geq \frac{1}{5}$. In this case, the overall signal space (of dim-$N$) are entirely occupied by the units with Pattern 1.2. This corresponds to a maximum DoF per user of $\frac{N}{5}$. We emphasize that $\frac{N}{5}$ is not necessarily an integer. We can use the technique of symbol extension to achieve a fractional DoF. For example, consider 5 channel uses. Then, the overall signal space is enlarged to be of dim-$5N$. Then, $N$ units with Pattern 1.2 can be constructed (provided that $M \geq \frac{N}{5}$), which achieves a total DoF of $6N$, or equivalently, a DoF per user per channel use of $\frac{N}{5}$. Similar symbol extension techniques will be used to achieve a fractional DoF without further notice. To summarize, the maximum DoF per user is given by $\min\left(M, \frac{N}{5}\right)$ with probability one when $\frac{1}{6} < \frac{M}{N} \leq \frac{1}{3}$.

*3) Case of $\frac{1}{3} < \frac{M}{N} < \frac{4}{9}$:* In this case, $\frac{M}{N}$ is sufficiently large to construct units with Pattern 1.3. Denote $\mathbf{H}_j = [\mathbf{H}_{j1}, \mathbf{H}_{j2}, \mathbf{H}_{j3}] \in \mathbb{C}^{N \times 3M}$, for $j = 1, 2$. From channel randomness, $\text{null}(\mathbf{H}_j)$ is of dimension $3M - N > 0$ with high probability. Thus, there exists a full column-rank matrix $\mathbf{U}_j \in \mathbb{C}^{3M \times (3M-N)}$ satisfying

$$\mathbf{H}_j\mathbf{U}_j = \mathbf{0}, \quad \text{for } j = 1, 2, \tag{17}$$

or equivalently

$$\sum_{k=1}^{K} \mathbf{H}_{jk}\mathbf{U}_{jk} = \mathbf{0}, \quad \text{for } j = 1, 2, \tag{18}$$

where $\mathbf{U}_j$ is partitioned as $\mathbf{U}_j = \left[\mathbf{U}_{j1}^T, \mathbf{U}_{j2}^T, \mathbf{U}_{j3}^T\right]^T$ with $\mathbf{U}_{jk} \in \mathbb{C}^{M \times (3M-N)}$, $\forall j, k$. From (18), the rank of $\mathbf{M} \in \mathbb{C}^{N \times 6(6M-N)}$ (defined in (4b)) is at most $\min\{N, 4(3M - N)\}$. From Lemma 10 in Appendix A and the fact that $\mathbf{H}_1$ and $\mathbf{H}_2$ are mutually independent, we see that $\mathbf{M}$ is of rank $\min\{N, 4(3M - N)\}$ with probability one.

Consider the case of $N > 4(3M - N)$, or equivalently, $\frac{M}{N} < \frac{5}{12}$. Then, $\text{rank}(\mathbf{M}) = 4(6M - N)$ with probability one. Denote by $\mathbf{u}_{jk}^{(l)}$ the $l$-th column of $\mathbf{U}_{jk}$. Then

$$\sum_{k=1}^{K} \mathbf{H}_{jk}\mathbf{u}_{jk}^{(l)} = \mathbf{0}, \text{ for } j = 1, 2, \text{ and } l = 1, \cdots, 3M - N. \tag{19}$$

[2]For $\frac{M}{N} > \frac{1}{6}$, there is enough freedom to construct units with Pattern 1.1 to occupy the overall signal space. Thus, units with Pattern 1.1 can be constructed to occupy any signal subspace left by the units with pattern 1.2. Similar arguments will be implicitly used throughout this paper.

Thus, $\text{span}(\{\mathbf{H}_{jk}\mathbf{u}_{jk}^{(l)}|\forall j,\forall k\})$ is of at most dim-4, for $l = 1,\cdots,3M-N$. Further, since $\text{rank}(\mathbf{M}) = 4(6M-N)$, we have $\dim(\text{span}(\{\mathbf{H}_{jk}\mathbf{u}_{jk}^{(l)}|\forall j,\forall k\})) = 4$ with probability one, for $l = 1,\cdots,3M-N$. Therefore, $\{\mathbf{H}_{jk}\mathbf{u}_{jk}^{(l)}|\forall j,k\}$ forms a unit with Pattern 1.3, for $l = 1,\cdots,3M-N$. In total, these $3M-N$ units with Pattern 1.3 occupy a dim-$4(3M-N)$ subspace with probability one. The remaining $N-4(3M-N)$ dimensions are for constructing $\frac{N-4(3M-N)}{5}$ units with Pattern 1.2. Thus, the achievable DoF per user is

$$d_{\text{user}} = (3M-N) + \frac{N-4(3M-N)}{5} = \frac{3M}{5}. \quad (20)$$

Now consider the remaining case of $\frac{5}{12} \le \frac{M}{N} < \frac{4}{9}$. In this case, the overall space (of dim-$N$) are all used for constructing units with Pattern 1.3. This corresponds to a maximum DoF per user of $\frac{N}{4}$. Thus, we conclude that the maximum DoF per user is given by $\min\left(\frac{3M}{5},\frac{N}{4}\right)$ with probability one when $\frac{1}{3} < \frac{M}{N} < \frac{4}{9}$.

*4) Case of $\frac{4}{9} \le \frac{M}{N} \le \frac{2}{3}$:* In this case, we will construct units to achieve $d_{\text{relay}} = 2$. Unlike the preceding cases in which signal alignment is done in a unit-by-unit manner, we will jointly align the signals of multiple units here, with the reason explained as follows. The unit-by-unit signal-alignment technique implies that the signals from different units can be distinguished at the relay (i.e., the relay is able to decode combinations of the signal streams in one unit without seeing any interference from the other units). However, this *decodability* is not necessarily required to accomplish clustered full data exchange. As a matter of fact, the relay is allowed to decode combinations of the signal streams belonging to different units, provided that these streams are from a same cluster. This means that joint signal alignment of multiple units can potentially outperform unit-by-unit based signal alignment, as seen in what follows.

We start with unit-by-unit signal alignment. To construct a unit with Pattern 1.4, we need to design beamforming vectors satisfying

$$\mathbf{H}_{j1}\mathbf{u}_{j1} + \mathbf{H}_{j2}\mathbf{u}_{j2} + \mathbf{H}_{j3}\mathbf{u}_{j3} = \mathbf{0}, \quad j = 1,2, \quad (21a)$$

$$\mathbf{H}_{11}\mathbf{u}_{11} + \mathbf{H}_{12}\mathbf{u}_{12} + \mathbf{H}_{21}\mathbf{u}_{21} + \mathbf{H}_{22}\mathbf{u}_{22} = \mathbf{0}, \quad (21b)$$

or equivalently, in a matrix form as

$$\begin{pmatrix} \mathbf{H}_{11} & \mathbf{H}_{12} & \mathbf{H}_{13} & \mathbf{0} & \mathbf{0} & \mathbf{0} \\ \mathbf{0} & \mathbf{0} & \mathbf{0} & \mathbf{H}_{21} & \mathbf{H}_{22} & \mathbf{H}_{23} \\ \mathbf{H}_{11} & \mathbf{H}_{12} & \mathbf{0} & \mathbf{H}_{21} & \mathbf{H}_{22} & \mathbf{0} \end{pmatrix}\mathbf{u} = \mathbf{B}\mathbf{u} = \mathbf{0} \quad (22)$$

where $\mathbf{u} = [\mathbf{u}_{11}^T,\mathbf{u}_{12}^T,\mathbf{u}_{13}^T,\mathbf{u}_{21}^T,\mathbf{u}_{22}^T,\mathbf{u}_{23}^T]^T$. As $\mathbf{B} \in \mathbb{C}^{3N\times 6M}$ is of full rank for sure, we require that $\frac{M}{N} > \frac{1}{2}$ to ensure that (22) has nontrivial solutions.

We next show that the above requirement on $\frac{M}{N}$ can be relaxed to $\frac{M}{N} \ge \frac{4}{9}$ when joint signal alignment of multiple units is considered. Our target is to construct full-rank beamforming matrices $\{\mathbf{U}_{jk} \in \mathbb{C}^{N\times t}\}$ satisfying

$$\mathbf{H}_{j1}\mathbf{U}_{j1} + \mathbf{H}_{j2}\mathbf{U}_{j2} + \mathbf{H}_{j3}\mathbf{U}_{j3} = \mathbf{0}, \quad j = 1,2, \quad (23a)$$

$$\sum_{j=1}^{2}\sum_{k=1}^{2}\mathbf{H}_{jk}\mathbf{U}_{jk}\widetilde{\mathbf{U}}_{jk} = \mathbf{0}, \quad (23b)$$

where $t$ represents the number of units involved, and $\widetilde{\mathbf{U}}_{jk} \in \mathbb{C}^{t\times t}$ is a full-rank matrix. We will determine $t$ and $\widetilde{\mathbf{U}}_{jk}$ shortly in this subsection. In the above, the existence of $\{\widetilde{\mathbf{U}}_{jk}\}$ implies that, when the relay decodes a combination from a certain cluster, signal streams from different units are allowed to interfere with each other, provided that these signal streams are from the same cluster. From (23), $\text{span}(\mathbf{M})$ (with $\mathbf{M}$ defined in (4b)) is of dimension $3t$; $\text{span}(\mathbf{M}_j)$ is of dimension $2t$. Then, $\text{null}(\mathbf{M}_{3-j}) \cap \text{span}(\mathbf{M})$ is of dimension $t$, and so the projection matrix $\mathbf{P}_j$ is of rank $t$. This implies that the relay can decode $t$ independent linear combinations of the messages from each cluster. As the received signal at the relay occupies a subspace of dim-$3t$, the above scheme achieves $d_{\text{relay}} = \frac{6t}{3t} = 2$ (which is equal to the $d_{\text{relay}}$ of Pattern 1.4).

We now describe how to construct $\{\mathbf{U}_{jk}\}$ satisfying (23). We show that, when $\frac{M}{N} = \frac{4}{9}$, this can be done with $t = \frac{N}{3}$. From the uplink/downlink symmetry, it suffices to only consider (23). From Lemma 11 in Appendix A, $\text{span}(\mathbf{H}_{j1},\mathbf{H}_{j2}) \cap \text{span}(\mathbf{H}_{j3})$ is of dimension $3M-N = \frac{N}{3}$ with probability one. Thus, there exist $\mathbf{U}_{jk} \in \mathbb{C}^{N\times\frac{N}{3}}$, for $j = 1,2$ and $k = 1,2,3$, satisfying (23a). Further, $\text{span}(\mathbf{H}_{j1}\mathbf{U}_{j1},\mathbf{H}_{j2}\mathbf{U}_{j2})$ is of dim-$\frac{2N}{3}$ with probability one. Noting that $\text{span}(\mathbf{H}_{11}\mathbf{U}_{11},\mathbf{H}_{12}\mathbf{U}_{12})$ and $\text{span}(\mathbf{H}_{21}\mathbf{U}_{21},\mathbf{H}_{22}\mathbf{U}_{22})$ are independent of each other, we obtain from Lemma 11 that $\dim(\text{span}(\mathbf{H}_{11}\mathbf{U}_{11},\mathbf{H}_{12}\mathbf{U}_{12}) \cap \text{span}(\mathbf{H}_{21}\mathbf{U}_{21},\mathbf{H}_{22}\mathbf{U}_{22})) = \frac{N}{3}$ with probability one. Let $\widetilde{\mathbf{H}} \in \mathbb{C}^{N\times\frac{N}{3}}$ give a basis of the intersection of $\text{span}(\mathbf{H}_{11}\mathbf{U}_{11},\mathbf{H}_{12}\mathbf{U}_{12}) \cap \text{span}(\mathbf{H}_{21}\mathbf{U}_{21},\mathbf{H}_{22}\mathbf{U}_{22})$. Then, since $\text{span}(\widetilde{\mathbf{H}}) \subset \text{span}(\mathbf{H}_{j1}\mathbf{U}_{j1},\mathbf{H}_{j2}\mathbf{U}_{j2})$, for $j = 1,2$, there exist $\widetilde{\mathbf{U}}_{jk} \in \mathbb{C}^{\frac{N}{3}\times\frac{N}{3}}$, for $j = 1,2$ and $k = 1,2$, satisfying

$$\begin{aligned} \widetilde{\mathbf{H}} &= \mathbf{H}_{11}\mathbf{U}_{11}\widetilde{\mathbf{U}}_{11} + \mathbf{H}_{12}\mathbf{U}_{12}\widetilde{\mathbf{U}}_{12} \\ &= -\mathbf{H}_{21}\mathbf{U}_{21}\widetilde{\mathbf{U}}_{21} - \mathbf{H}_{22}\mathbf{U}_{22}\widetilde{\mathbf{U}}_{22}. \end{aligned} \quad (24)$$

Then, (23b) is also met. Therefore, we conclude that a per-user DoF of $\frac{N}{3}$ is achievable when $\frac{4}{9} \le \frac{M}{N} \le \frac{2}{3}$.

*5) Case of $\frac{M}{N} > \frac{2}{3}$:* In this case, $\frac{M}{N}$ is large enough to construct units with Pattern 1.5. Specifically, the intersection of three subspaces $\text{span}(\mathbf{H}_{jk})$, $k = 1,2,3$, is of dim-$(3M-2N)$ with probability one. That is, with high probability, there exist unitary matrices $\mathbf{U}_{jk} \in \mathbb{C}^{M\times(3M-2N)}$ for $\forall j,k$ satisfying

$$\mathbf{H}_{j1}\mathbf{U}_{j1} = \mathbf{H}_{j2}\mathbf{U}_{j2} = \mathbf{H}_{j3}\mathbf{U}_{j3}, \quad j = 1,2. \quad (25)$$

From (25), $\mathbf{M} \in \mathbb{C}^{N\times 6(3M-2N)}$ is of rank $\min(N, 2(3M-2N))$ with probability one.

Now suppose $N > 2(3M-2N)$, or equivalently, $\frac{M}{N} < \frac{5}{6}$. Then, $\mathbf{M}$ is of rank $2(3M-2N)$ with probability one. Denote by $\mathbf{u}_{jk}^{(l)}$ the $l$-th column of $\mathbf{U}_{jk}$. From (25), $\{\mathbf{H}_{jk}\mathbf{u}_{jk}^{(l)}|j = 1,2; k = 1,2,3\}$ form a unit with Pattern 1.5, and in total, there are $3M-2N$ of such units, occupying a signal subspace of dim-$(2(3M-2N))$. The remaining signal space of dim-$(N-2(3M-2N))$ is used to support Pattern 1.4. Thus

$$d_{\text{user}} = (3M-2N) + \frac{N-2(3M-2N)}{3} = M - \frac{N}{3}. \quad (26)$$

We now consider the remaining case of $\frac{M}{N} \ge \frac{5}{6}$. The overall signal space is wholly occupied by units with Pattern 1.5. This corresponds to a maximum DoF per user of $\frac{N}{2}$. Therefore, the maximum DoF per user is given by $\min\left(M - \frac{N}{3}, \frac{N}{2}\right)$, which concludes the proof of Lemma 1.

## IV. Generalization to Arbitrary $L$ and $K$

In this section, we generalize the DoF results to an arbitrary network configuration of $L$ and $K$. We start with some notions arising from the special case studied in the preceding section.

### A. Definitions

We define a corner point of an achievable DoF curve as follows.

*Definition 1:* Given an achievable DoF curve of $d_{\text{user}} = \tilde{f}(M, N)$ with $N = N_0$, a point $(\frac{M}{N_0}, d_{\text{user}})$ is said to be achievable if $d_{\text{user}} \leq \tilde{f}(M, N_0)$.

*Definition 2:* (*Corner Point*) Given an achievable DoF curve of $d_{\text{user}} = \tilde{f}(M, N)$ with $N = N_0$, an achievable point $(\frac{M_0}{N_0}, d_0)$ is said to be a *corner point* if $(\frac{M}{N_0}, \frac{d_0 M}{M_0})$ is not achievable for any $M > M_0$.

A corner point is achieved when the overall relay's signal space is occupied by units following a *single* pattern. For the case of $L = 2$ and $K = 3$, five patterns are considered in deriving the achievable DoF in Lemma 1, and therefore, there are five *candidate* corner points, i.e., $(\frac{1}{6}, \frac{N}{6})$, $(\frac{1}{5}, \frac{N}{5})$, $(\frac{5}{12}, \frac{N}{4})$, $(\frac{4}{9}, \frac{N}{3})$, and $(\frac{5}{6}, \frac{N}{2})$, corresponding to Patterns 1.1 to 1.5, respectively. Some of them (namely, $(\frac{1}{5}, \frac{N}{5})$, $(\frac{4}{9}, \frac{N}{3})$, and $(\frac{5}{6}, \frac{N}{2})$) are indeed corner points of the DoF curve given in Theorem 3 (see Fig. 3); the others (namely, $(\frac{1}{6}, \frac{N}{6})$ and $(\frac{5}{12}, \frac{N}{4})$) are not as they are obscured by the *real* corner points.

To obtain the achievable DoF in Theorem 3, it suffices for us to identify the corner points and then to apply the antenna-disablement lemma to these corner points. To be specific, define the $g$-function as

$$g_{(a,b)}(x) = \begin{cases} \frac{bx}{a}, & \text{for } x < a \\ b, & \text{for } x \geq a \end{cases} \tag{27}$$

where $a$ and $b$ are constant coefficients. Then, we can represent the achievable DoF in Theorem 3 using corner points as

$$d_{\text{user}} = \max\left(g_{(\frac{1}{5}, \frac{N}{5})}\left(\tfrac{M}{N}\right), g_{(\frac{4}{9}, \frac{N}{3})}\left(\tfrac{M}{N}\right), g_{(\frac{5}{6}, \frac{N}{2})}\left(\tfrac{M}{N}\right)\right). \tag{28}$$

We categorize all corner points into two types, namely, Type-I and Type-II, based on the value of the antenna ratio $\frac{M}{N}$. Type-I corner points are those that fall into the range of $\frac{M}{N} \leq \frac{1}{K}$. In this $\frac{M}{N}$ range, we have $KM \leq N$. Then the nullspace of $\mathbf{H}_j = [\mathbf{H}_{j1}, \mathbf{H}_{j2}, \cdots, \mathbf{H}_{jK}] \in \mathbb{C}^{N \times KM}$ contains no vector except the trivial zero vector, implying that, no matter how a pattern is constructed, the $K$ channel vectors of a common cluster $j$ always span a subspace of dim-$K$. Therefore, to construct patterns corresponding to Type-I corner points, we only need to consider signal space alignment between clusters. For the case of $L = 2$ and $K = 3$, $(\frac{1}{5}, \frac{N}{5})$ is a Type-I corner point, as seen from Fig. 3.

Type-II corner points are those that fall into the range of $\frac{M}{N} > \frac{1}{K}$. In this $\frac{M}{N}$ range, we have $KM > N$. Thus, there are nontrivial vectors in the null space of $[\mathbf{H}_{j1}, \mathbf{H}_{j2}, \cdots, \mathbf{H}_{jK}]$, implying that we need to consider signal space alignment not only between clusters, but also within each cluster. For the case of $L = 2$ and $K = 3$, Type-II corner points include $(\frac{4}{9}, \frac{N}{3})$ and $(\frac{5}{6}, \frac{N}{2})$, as seen from Fig. 3.

To generalize the DoF results to the case of arbitrary $L$ and $K$, we need to systematically identify all possible corner points (or equivalently, the corresponding patterns), as elaborated in what follows.

### B. Main Result

We consider the construction of Type-I candidate corner points for a general setup of $L$ and $K$. The results are presented below, with the proofs given in Subsections IV-C and IV-D.

*Lemma 4:* For the considered MIMO mRC with $L$ clusters and $K$ users per cluster and $\frac{M}{N} \in (\frac{1}{2K}, \frac{1}{K}]$, the following candidate corner points $(\frac{M}{N}, d_{\text{user}})$ are achievable:

$$\frac{M}{N} = \max_{l': 2 \leq l' \leq l} \frac{1}{l'K}\left(l' - 1 + \frac{\binom{l}{l'}}{\binom{L}{l'}(lK - l + 1)}\right) \tag{29a}$$

$$d_{\text{user}} = \frac{N}{lK - l + 1} \cdot \frac{l}{L}, \quad \text{for } l = 2, \cdots, L. \tag{29b}$$

*Lemma 5:* For the considered MIMO mRC with $L$ clusters and $K$ users per cluster and $\frac{M}{N} \in (0, \frac{1}{2K}]$, the following candidate corner points $(\frac{M}{N}, d_{\text{user}})$ are achievable:

$$\frac{M}{N} = \frac{1}{(tl+1)K}\left(l + \frac{1}{(tl+1)K - l}\right) \tag{30a}$$

$$d_{\text{user}} = \frac{N}{(tl+1)K - l} \cdot \frac{tl+1}{L} \tag{30b}$$

where $l = \left\lfloor \frac{L-1}{t} \right\rfloor$, and $t = 2, \cdots, L-1$.

We now consider the construction of Type-II corner points, with the result given below. The proof can be found in Subsection IV-E.

*Lemma 6:* For the considered MIMO mRC with $L$ clusters and $K$ users per cluster $\frac{M}{N} > \frac{1}{K}$, the following candidate corner points $(\frac{M}{N}, d_{\text{user}})$ are achievable:

$$\frac{M}{N} = \frac{1}{K}\left(K - k + \frac{1}{k(l-1)+1}\right) \tag{31a}$$

$$d_{\text{user}} = \frac{N}{k(l-1)+1} \cdot \frac{l}{L} \tag{31b}$$

where $k = 1, \cdots, K-1$, and $l = 2, \cdots, L$.

With Lemmas 4-6, we present an achievable DoF of the MIMO mRC with an arbitrary setup of $(L, K, M, N)$ as follows. Denote by $\mathcal{I}$ the set of all the candidate corner points $(\frac{M}{N}, d_{\text{user}})$ specified in (29), (30), and (31). Then, we obtain the following main theorem of this paper.

*Theorem 7:* For the considered $M$-by-$N$ MIMO mRC with $L$ clusters and $K$ users per cluster, the following DoF per user is achievable:

$$d_{\text{user}} = \max_{(a,b) \in \mathcal{I}} g_{(a,b)}\left(\tfrac{M}{N}\right), \tag{32}$$

where the $g$-function is defined in (27).

*Remark 1:* Theorem 3 is a special case of Theorem 7. Specifically, for $L = 2$ and $K = 3$, we readily identify the following visible corner points given in Lemmas 4 and 5: Type-I corner point $(\frac{M}{N}, d_{\text{user}}) = (\frac{1}{5}, \frac{N}{5})$ given by (29) with $l=2$; Type-II corner point $(\frac{4}{9}, \frac{N}{3})$ given by (31) with $l=2$ and $k=2$; and Type-II corner point $(\frac{5}{6}, \frac{N}{2})$ given by (31) with $l=2$ and $k=1$. Then, Theorem 3 follows by (32).

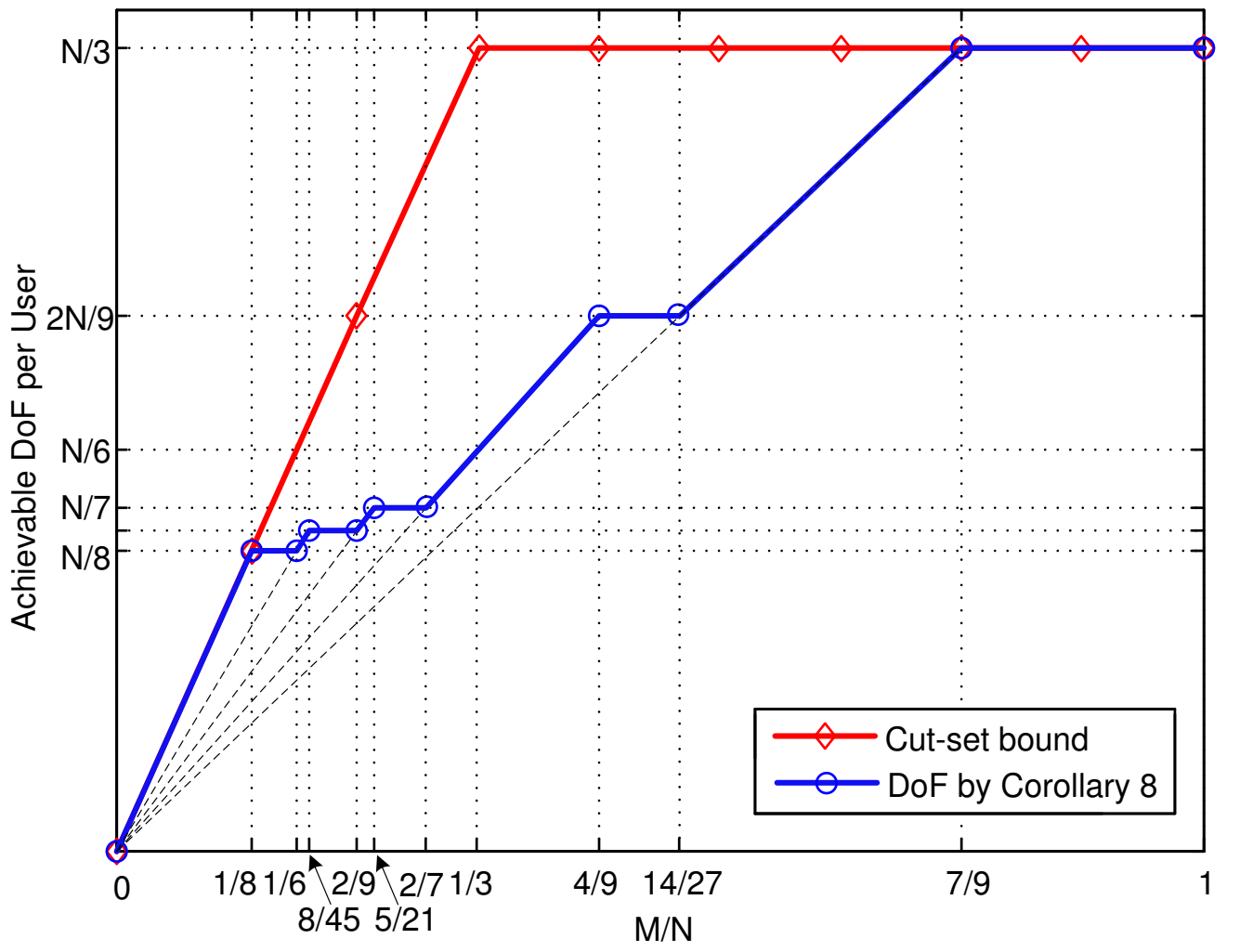

Fig. 4. The achievable DoF per user and the DoF cut-set bound with respect to the antenna ratio $\frac{M}{N}$ for the 3-cluster MIMO mRC operating in the clustered full data exchange model. Each cluster has 3 users.

Letting $L=3$ and $K=3$ in Theorem 7, we obtain the following result.

*Corollary 8:* For the considered $M$-by-$N$ MIMO mRC with $L=3$ and $K=3$, an achievable DoF per user is

$$d_{\text{user}} = \begin{cases} \min\left(M, \dfrac{N}{8}\right), & \dfrac{M}{N} \le \dfrac{1}{6} \\ \min\left(\dfrac{3M}{4}, \dfrac{2N}{15}\right), & \dfrac{1}{6} < \dfrac{M}{N} \le \dfrac{2}{9} \\ \min\left(\dfrac{3M}{5}, \dfrac{N}{7}\right), & \dfrac{2}{9} < \dfrac{M}{N} \le \dfrac{2}{7} \\ \min\left(\dfrac{M}{2}, \dfrac{N}{6}\right), & \dfrac{2}{7} < \dfrac{M}{N} \le \dfrac{14}{27} \\ \min\left(\dfrac{3}{7}M, \dfrac{N}{3}\right), & \dfrac{M}{N} > \dfrac{14}{27}. \end{cases} \quad (33)$$

*Proof:* For $L=3$ and $K=3$, the following corner points are identified: $(\frac{1}{8}, \frac{N}{8})$ given by (30) with $t=2$; $(\frac{8}{45}, \frac{2N}{15})$ given by (29) with $l=2$; $(\frac{5}{21}, \frac{N}{7})$ given by (29) with $l=3$; $(\frac{4}{9}, \frac{2N}{9})$ given by (31) with $l=2$ and $k=2$; and $(\frac{7}{9}, \frac{N}{3})$ given by (31) with $l=3$ and $k=1$. Then, the corollary follows by (32). ■

*Remark 2:* The achievable DoF against $\frac{M}{N}$ given by Corollary 8 is plotted in Fig. 4. For comparison, we also provide the cut-set outer bound in (11). From Fig. 4, the achievable bound matches the cut-set outer bound when $\frac{M}{N} \ge \frac{7}{9}$ and $\frac{M}{N} \le \frac{1}{8}$, implying that the DoF capacity of the consider MIMO mRC is achievable in these ranges of $\frac{M}{N}$. This result can be generalized to the case of arbitrary $L$ and $K$, as detailed below.

*Corollary 9:* For the considered $M$-by-$N$ MIMO mRC with $L$ clusters and $K$ users in each cluster, the DoF capacity of $d_{\text{user}} = M$ is achievable when $\frac{M}{N} \in \left(0, \frac{1}{LK-1}\right]$; and the DoF capacity of $d_{\text{user}} = \frac{N}{L}$ is achievable when $\frac{M}{N} \in \left[\frac{(K-1)L+1}{KL}, \infty\right)$.

*Proof:* For $\frac{M}{N} \in \left(0, \frac{1}{LK-1}\right]$, $(\frac{1}{LK-1}, M)$ is a corner point given by (30) with $l=1$ and $t=1$, which achieves the DoF outer bound in (11); for $\frac{M}{N} \in \left[\frac{(K-1)L+1}{KL}, \infty\right)$, $(\frac{(K-1)L+1}{KL}, \frac{N}{L})$ is a corner point given by (31) with $l=L$ and $k=1$, which again achieves the DoF outer bound. ■

*Remark 3:* We see from Corollary 9 that $d_{\text{sum}} = KLd_{\text{user}}$ is proportional to $K$ (i.e., the number of users per cluster), when $\frac{M}{N} \in \left(0, \frac{1}{LK-1}\right]$ and $\frac{M}{N} \in \left[\frac{(K-1)L+1}{KL}, \infty\right)$. It can be verified that, with the achievable DoF given in Theorem 7, the linearity of $d_{\text{sum}}$ in $K$ holds for any $(M, N)$ configurations. Intuitively, this linearity is due to the fact that: in clustered full data exchange, each relay antenna may transmit combinations of $K$ messages in a cluster (one from each user in the cluster) in $K-1$ time slots; each user in the cluster receives $K-1$ combinations, and is able to decode the $K-1$ messages from the other users by utilizing the self-message, which results in $K$ DoF per relay antenna, or $KN$ DoF in total. Note that this linearity of DoF in $K$ in general does not hold for other data exchange models. For example, for pairwise data exchange in [17], the total DoF of the MIMO mRC is bounded by $2N$ which does not scale with the number of users.

## C. Proof of Lemma 4

Without loss of generality, we assume that the pattern to be constructed involves $l \in [2, L]$ active clusters, and index these clusters as $j_{[1]}, j_{[2]}, \cdots, j_{[l]}$ in the ascending order. Denote $\mathbf{H}_j = [\mathbf{H}_{j1}, \mathbf{H}_{j2}, \cdots, \mathbf{H}_{jK}]$. Our target is to construct beamforming vectors $\{\mathbf{u}_{j_{[1]}}, \cdots, \mathbf{u}_{j_{[l]}}\}$ satisfying the following equation:

$$\begin{pmatrix} \mathbf{H}_{j_{[1]}} & \mathbf{H}_{j_{[2]}} & & & \\ & \mathbf{H}_{j_{[2]}} & \mathbf{H}_{j_{[3]}} & & \\ & & \ddots & \ddots & \\ & & & \mathbf{H}_{j_{[l-1]}} & \mathbf{H}_{j_{[l]}} \end{pmatrix} \begin{pmatrix} \mathbf{u}_{j_{[1]}} \\ \mathbf{u}_{j_{[2]}} \\ \vdots \\ \mathbf{u}_{j_{[l]}} \end{pmatrix} = \mathbf{B}_1 \mathbf{u} = \mathbf{0}. \quad (34)$$

In the above, the block matrix $\mathbf{B}_1$ contains $l-1$ block-rows. For each index $i$, the $i$-th block-row of $\mathbf{B}_1$ corresponds to a vector in $\text{null}([\mathbf{H}_{j_{[i]}}, \mathbf{H}_{j_{[i+1]}}])$. The existence of such vectors is guaranteed by the fact that, for $\frac{M}{N} \in (\frac{1}{2K}, \frac{1}{K}]$, $\text{null}([\mathbf{H}_j, \mathbf{H}_{j'}])$, for $\forall j, j'$, is of dimension $2KM - N \ge 0$ for sure.

The maximum number of block-rows of $\mathbf{B}_1$ is limited to $L-1$, since it is required that every new block-row added into $\mathbf{B}_1$ must contain one and only one new $\mathbf{H}_j$.[3] Also, each unit following the pattern in (34) contains $lK$ spatial streams, with one from each of the users in the $l$ active clusters; the subspace occupied by each unit is of dimension $lK - l + 1$, as $\mathbf{B}_1$ in (34) contains $l-1$ block-rows. Furthermore, for each active cluster $j_{[t]}$, the interference consists of the $(l-1)K$

[3]Otherwise, there must exist an $\mathbf{H}_j$ repeated in $\mathbf{B}_1$ at different columns. For example, suppose $\mathbf{H}_{j_{[3]}}$ is replaced by $\mathbf{H}_{j_{[1]}}$ in (34). Then, subtracting the first block-row of $\mathbf{B}_1$ by the second block-row, we obtain $\mathbf{H}_{j_{[1]}}\mathbf{U}_{j_{[1]}} - \mathbf{H}_{j_{[1]}}\mathbf{U}_{j_{[3]}} = \mathbf{0}$, or equivalently, $\mathbf{H}_{j_{[1]}}(\mathbf{U}_{j_{[1]}} - \mathbf{U}_{j_{[3]}}) = \mathbf{0}$. Note that $\mathbf{U}_{j_{[1]}} \neq \mathbf{U}_{j_{[3]}}$, as they correspond to different spatial streams in a pattern. Therefore, $\mathbf{H}_{j_{[1]}}(\mathbf{U}_{j_{[1]}} - \mathbf{U}_{j_{[3]}}) = \mathbf{0}$ implies that the nullspace of $\mathbf{H}_{j_{[1]}} \in C^{N \times MK}$ is not trivial, i.e., $MK > N$. This contradicts with the condition of $\frac{M}{N} \in (\frac{1}{2K}, \frac{1}{K}]$ in Lemma 4.

spatial streams from the other $l-1$ clusters, satisfying the following equation:

$$\begin{pmatrix} \ddots & \ddots & & \\ & \mathbf{H}_{j_{[t-2]}} & \mathbf{H}_{j_{[t-1]}} & \\ & & \mathbf{H}_{j_{[t-1]}} & \mathbf{H}_{j_{[t+1]}} \\ & & & \ddots & \ddots \end{pmatrix} \begin{pmatrix} \vdots \\ \mathbf{u}_{j_{[t-1]}} \\ -\mathbf{u}_{j_{[t+1]}} \\ \vdots \end{pmatrix} = \mathbf{0}. \quad (35)$$

The above matrix has $l-2$ block-rows and is of full row rank. Then, the interference of cluster $j_t$ spans a subspace of dimension $(l-1)K-(l-2)=(l-1)(K-1)+1$. Recall that the total signal space spanned by the pattern (34) is of dimension $l(K-1)+1$. Therefore, the relay sees an interference-free subspace of dimension $K-1 \geq 1$, and can obtain one combination of the $K$ signal streams from cluster $j_t$ without interference for sure. As the cluster index $j_t$ is arbitrarily chosen, we conclude that a unit following pattern (34) is decodable for sure at the relay (as well as at the user ends due to the uplink/downlink symmetry).

From Lemma 12, the number of independent units can be constructed following pattern (34) (with a given selection of the index set $\{j_{[1]}, j_{[2]}, \cdots, j_{[l]}\}$) is given by the dimension of the nullspace of $\mathbf{B}_1$. However, by considering all possible index selections, there are a family of $\binom{L}{l}$ different patterns in the form of (34). We need to determine how many linearly independent units following this pattern family can be constructed, as detailed below.

First, from the channel randomness, $\mathbf{B}_1$ in (34) is of full row rank with probability one, and thus the nullspace of $\mathbf{B}_1$ is of dimension $lMK-(l-1)N$ for sure. Then, $lMK-(l-1)N$ linearly independent units can be constructed for any given selection of the index set $\{j_{[1]}, j_{[2]}, \cdots, j_{[l]}\}$, as ensured by Lemma 10. Noting all the $\binom{L}{l}$ different ways to select the index set, we see that the total number of units is limited by

$$c_l = \binom{L}{l}(lMK-(l-1)N). \quad (36)$$

Second, each block row of $\mathbf{B}_1$ in (34) corresponds to an equation of the form $\mathbf{H}_j\mathbf{u}_j + \mathbf{H}_{j'}\mathbf{u}_{j'} = \mathbf{0}$ with $j \neq j'$. For any given index pair $(j, j')$ with $j \neq j'$, $\mathrm{null}([\mathbf{H}_j, \mathbf{H}_{j'}])$ has $2MK-N$ linearly independent nullspace vectors of the form $[\mathbf{u}_j^T, \mathbf{u}_{j'}^T]^T$ for sure. This means, the equation $\mathbf{H}_j\mathbf{u}_j + \mathbf{H}_{j'}\mathbf{u}_{j'} = \mathbf{0}$ cannot appear in all the constructed units for more than $2MK-N$ times. This gives another limitation on the number of units that can be constructed. Specifically, there are in total $\binom{L}{l}$ different ways to select $l$ active clusters from $L$ clusters; there are $\binom{L-2}{l-2}$ of these selections in which both clusters $j$ and $j'$ are selected. Let $c_2$ be the total number of units constructed. Then, the number of $\mathbf{H}_j\mathbf{u}_j + \mathbf{H}_{j'}\mathbf{u}_{j'} = \mathbf{0}$ used in constructing $c_2$ units is given by

$$c_2 \cdot \frac{\binom{L-2}{l-2}}{\binom{L}{l}} = c_2 \cdot \frac{\binom{l}{2}}{\binom{L}{2}}. \quad (37)$$

The above number cannot exceed $2MK-N$ (i.e., the number of independent vectors in $\mathrm{null}([\mathbf{H}_j, \mathbf{H}_{j'}])$). Therefore, the total number of units following pattern (34) is limited by

$$c_2 = \frac{\binom{L}{2}(2MK-N)}{\binom{l}{2}} \quad (38)$$

Third, there are more constraints on constructing units with the pattern in (34). Let $j'_{[1]}, j'_{[2]}, \cdots, j'_{[l']}$ (without loss of generality, in the ascending order) be a set of $l'(\leq l)$ distinct indexes selected from $\{1, \cdots, L\}$. Consider the following sub-pattern (similar to (34)) as

$$\begin{pmatrix} \mathbf{H}_{j'_{[1]}} & \mathbf{H}_{j'_{[2]}} & & & \\ & \mathbf{H}_{j'_{[2]}} & \mathbf{H}_{j'_{[3]}} & & \\ & & \ddots & \ddots & \\ & & & \mathbf{H}_{j'_{[l'-1]}} & \mathbf{H}_{j'_{[l']}} \end{pmatrix} \mathbf{u}' = \mathbf{B}'_1\mathbf{u}' = \mathbf{0}. \quad (39)$$

A key observation is that every block-row of $\mathbf{B}'_1$ can be obtained by linearly combining the block-rows of $\mathbf{B}_1$, provided that $\{j'_{[1]}, j'_{[2]}, \cdots, j'_{[l']}\} \subseteq \{j_{[1]}, j_{[2]}, \cdots, j_{[l]}\}$. To see this, it suffices for us to only consider the first block row of $\mathbf{B}'_1$. As $\{j'_{[1]}, j'_{[2]}\} \subseteq \{j_{[1]}, j_{[2]}, \cdots, j_{[l]}\}$, we have $j'_{[1]} = j_{[p]}$ and $j'_{[2]} = j_{[p']}$ for certain integers $p$ and $p'$ with $p < p'$. Consider the following combination of the $p$-th to $(p'-1)$-th block-rows of $\mathbf{B}_1$: $\mathrm{row}_p - \mathrm{row}_{p+1} + \mathrm{row}_{p+2} + \cdots + (-1)^{p'-p-1} \times \mathrm{row}_{p'-1}$. This yields $\mathbf{H}_{j_{[p]}}\mathbf{u}_{j_{[p]}} + (-1)^{p'-p-1}\mathbf{H}_{j_{[p']}}\mathbf{u}_{j_{[p']}} = \mathbf{0}$, or equivalently, $[\mathbf{H}_{j_{[p]}} \ \mathbf{H}_{j_{[p']}}][\mathbf{u}^T_{j_{[p]}} \ (-1)^{p'-p-1}\mathbf{u}^T_{j_{[p']}}]^T = \mathbf{0}$, which corresponds to the first block-row of $\mathbf{B}'_1$.

With the above observation, we see that every vector $\mathbf{u}$ satisfying (34) corresponds to a unique vector $\mathbf{u}'$ satisfying the sub-pattern in (39), provided that $\{j'_{[1]}, j'_{[2]}, \cdots, j'_{[l']}\} \subseteq \{j_{[1]}, j_{[2]}, \cdots, j_{[l]}\}$. This means, for all the linearly independent units following pattern (34), the sub-pattern in (39) cannot be used for more than $l'MK-(l'-1)N$ times (as the dimension of $\mathrm{null}(\mathbf{B}'_1)$ is $l'MK-(l'-1)N$ for sure). The chance of a random selection of $\{j'_{[1]}, j'_{[2]}, \cdots, j'_{[l']}\}$ satisfying $\{j'_{[1]}, j'_{[2]}, \cdots, j'_{[l']}\} \subseteq \{j_{[1]}, j_{[2]}, \cdots, j_{[l]}\}$ is given by

$$\frac{\binom{L-l'}{l-l'}}{\binom{L}{l}} = \frac{\binom{l}{l'}}{\binom{L}{l'}}. \quad (40)$$

Therefore, the total number of linearly independent units following (34) cannot exceed

$$c_{l'} = \frac{\binom{L}{l'}(l'MK-(l'-1)N)}{\binom{l}{l'}}, \text{ for } l' = 2, 3, \cdots, l. \quad (41)$$

Note that (36) and (38) are two special cases of (41) by letting $l' = l$ and $l' = 2$, respectively.

Fourth, besides those in the form of (39), we shall consider more sub-patterns to guarantee that linearly independent units are constructed, as detailed below. Let $\mathbf{B}''_1$ be a block matrix with each block-row given by an arbitrary combination of the block-rows of $\mathbf{B}_1$ in (34). We require that the nullspace of $\mathbf{B}''_1$ contains enough linearly independent vectors to support the linearly independent units constructed. Without loss of generality, we assume that the sub-pattern $\mathbf{B}''_1$ involves $l'$ clusters, and index these clusters as $\{j'_{[1]}, j'_{[2]}, \cdots, j'_{[l']}\}$. For $\{j'_{[1]}, j'_{[2]}, \cdots, j'_{[l']}\}$, we construct $\mathbf{B}'_1$ in the form of (39). Further, we have the following facts: first, every block-row of $\mathbf{B}''_1$ can be expressed as a linear combination of the block-rows of $\mathbf{B}'_1$, implying that the rank of $\mathbf{B}''_1$ is no greater than the rank of $\mathbf{B}'_1$, or equivalently, from the rank-nullity theorem, the dimension of $\mathrm{null}(\mathbf{B}''_1)$ is no less than the dimension of

null($\mathbf{B}_1'$); second, sub-patterns $\mathbf{B}_1'$ and $\mathbf{B}_1''$ have the same times of occurrence in the pattern family of (34) (by noting the fact that $\mathbf{B}_1'$ and $\mathbf{B}_1''$ involve the same set of active clusters). Therefore, provided that the requirement for sub-patterns of the form $\mathbf{B}_1'$ is met, the nullspace requirement for the sub-patterns of the form $\mathbf{B}_1''$ is automatically met.

Based on all the above discussions, we conclude that the maximum number of linearly independent units following pattern (34) is given by $c_{\min} = \min_{2\le l'\le l} c_{l'}$. To achieve a corner point, these $c_{\min}$ units span the overall relay's signal space of dimension $N$. Thus, we have $c_{\min}(lK-l+1)=N$, yielding $\frac{M}{N}$ in (29a). Also, as each unit spans a subspace of dimension $lK-l+1$, the corresponding $d_{\text{user}}$ is given by $d_{\text{user}} = \frac{N}{lK-l+1}\cdot\frac{l}{L}$, which concludes the proof.

### D. Proof of Lemma 5

Let $t$ be an integer in $[2, L-1]$, and $l = \left\lfloor \frac{L-1}{t} \right\rfloor$. For $\frac{M}{N} \in \left(\frac{1}{(t+1)K}, \frac{1}{tK}\right]$, we construct a pattern involving $tl+1$ active clusters, indexed as $j_{[1]}, j_{[2]}, \cdots, j_{[tl+1]}$, with the pattern matrix given below:

$$\mathbf{B}_2 = \begin{pmatrix} \mathbf{H}_{j_{[1]}} \cdots \mathbf{H}_{j_{[t+1]}} & & & \\ & \mathbf{H}_{j_{[t+1]}} \cdots \mathbf{H}_{j_{[2t+1]}} & & \\ & & \ddots \quad \ddots & \\ & & & \mathbf{H}_{j_{[t(l-1)+1]}} \cdots \mathbf{H}_{j_{[tl+1]}} \end{pmatrix}. \tag{42}$$

In the above, the block matrix $\mathbf{B}_2$ has $l$ block-rows; every block-row contains $t+1$ nonzero blocks; every new block-row added into $\mathbf{B}_2$ contains $t$ new $\mathbf{H}_j$s until the unused $\mathbf{H}_j$s are not enough to construct a new block-row, which implies $l = \left\lfloor \frac{L-1}{t} \right\rfloor$.

We first verify that (42) defines a valid pattern. Clearly, each unit following (42) contains $(tl+1)K$ spatial streams; the subspace spanned by each unit is of dimension $(tl+1)K-l$. For any index $j$, the interference of cluster $j$ contains $tlK$ spatial streams which span a subspace of dimension $tlK-(l-1)$. Therefore, for each cluster $j$, the relay sees an interference-free subspace of dimension $K-1\ge 1$, and can obtain one combination of the $K$ signal streams from cluster $j$ for sure. Thus, a unit with pattern (34) is decodable at the relay (as well as at the users due to the uplink/downlink symmetry).

We next determine the number of linearly independent units that can be constructed with pattern (42). From Lemma 12, the maximum number of linearly independent units allowed is given by the nullspace dimension of the pattern matrix. Note that null($\mathbf{B}_2$) is of dimension $(tl+1)MK-lN$ with probability one. Therefore, we are able to construct $\tilde{c} = ((tl+1)MK-lN$ units following pattern (42). [4] To achieve a corner point, these $\tilde{c}$ units span the overall relay's signal space of dimension $N$. Thus, we have $\tilde{c}((tl+1)K-l) = N$, yielding $\frac{M}{N}$ given in (30a). Also, as each unit is of dimension $(tl+1)K-l$, the corresponding $d_{\text{user}}$ is given by $d_{\text{user}} = \frac{N}{(tl+1)K-l}\cdot\frac{tl+1}{L}$, which concludes the proof.

[4] We also need to guarantee that the nullspace of any sub-pattern of $\mathbf{B}_2'\mathbf{u}'=\mathbf{0}$ (with each block-row of the block matrix $\mathbf{B}_2'$ formed by a linear combination of the block-rows of $\mathbf{B}_2$) is not overcrowded to support $\tilde{c}$ units. From linear algebra, the rank of $\mathbf{B}_2'$ is no greater than the rank of $\mathbf{B}_2$. Thus, from the rank-nullity theorem, the dimension of null($\mathbf{B}_2'$) is no less than that of null($\mathbf{B}_2$), implying that null($\mathbf{B}_2'$) is not the bottleneck to support linearly independent units. Thus, the maximum number of units following pattern (42) is limited by $\tilde{c}$.

### E. Proof of Lemma 6

We generalize the construction of Pattern 1.4 in Section III as follows. Let $l$ be the number of active clusters, $m$ be the number of spatial streams transmitted by each active user, and $k$ be an arbitrary integer $\in [1, K]$. For the uplink channel, we design the beamforming matrices $\{\mathbf{U}_{jk}\in\mathbb{C}^{M\times m}\}$ to satisfy

$$\mathbf{H}_{j1}\mathbf{U}_{j1} + \cdots + \mathbf{H}_{jk}\mathbf{U}_{jk} + \mathbf{H}_{j(k+t)}\mathbf{U}_{j(k+t)} = \mathbf{0}, \tag{43}$$

for $j=1,\cdots,l,$ and $t=1,\cdots,K-k$, or equivalently in a matrix form as

$$\begin{pmatrix} \mathbf{H}_{j1} \cdots \mathbf{H}_{jk} & \mathbf{H}_{j(k+1)} & & \mathbf{0} \\ \vdots \quad \ddots \quad \vdots & & \ddots & \\ \mathbf{H}_{j1} \cdots \mathbf{H}_{jk} & \mathbf{0} & & \mathbf{H}_{jK} \end{pmatrix} \begin{pmatrix} \mathbf{U}_{j1} \\ \vdots \\ \mathbf{U}_{jK} \end{pmatrix} = \mathbf{B}_3\mathbf{U}_j = \mathbf{0}. \tag{44}$$

for $j = 1,\cdots,l$. In the above, (43) means that span($\mathbf{H}_{j(k+t)}\mathbf{U}_{j(k+t)}$), for $t=1,\cdots,K-k$, fall into span($\mathbf{H}_{j1}\mathbf{U}_{j1},\cdots,\mathbf{H}_{jk}\mathbf{U}_{jk}$), or equivalently, span($\mathbf{H}_{j1}\mathbf{U}_{j1},\cdots,\mathbf{H}_{jK}\mathbf{U}_{jK}$) = span($\mathbf{H}_{j1}\mathbf{U}_{j1},\cdots,\mathbf{H}_{jk}\mathbf{U}_{jk}$) is of dimension $mk$. From Lemma 12, to ensure that $m$ linearly independent units with pattern in (43) can be constructed, it is required that the nullspace of $\mathbf{B}_3\in\mathbb{C}^{(K-k)N\times KM}$ in (43) is of dimension $m$. From channel randomness, this can be met when

$$KM-(K-k)N = m. \tag{45}$$

On the other hand, to ensure that the relay can decode $m$ linear equations for each cluster $j$, the dimension of the interference space (spanned by the signals from the other $l-1$ active clusters) cannot exceed $N-mk(L-1)$, by noting that the subspace spanned by the interference from each cluster $j'\ne j$, denoted by span($\mathbf{H}_{j'1}\mathbf{U}_{j'1},\cdots,\mathbf{H}_{j'K}\mathbf{U}_{j'K}$), is of dimension $mk$. That is,

$$N-mk(l-1) = m. \tag{46}$$

Combining (45) and (46), we obtain

$$m = \frac{N}{k(l-1)+1} \text{ and } \frac{M}{N} = \frac{1}{K}\left(K-k+\frac{1}{k(l-1)+1}\right). \tag{47}$$

Noting that only $l$ of $L$ clusters are active, we obtain

$$d_{\text{user}} = m\cdot\frac{l}{L} = \frac{N}{k(l-1)+1}\cdot\frac{l}{L}. \tag{48}$$

Considering $l=2,\cdots,L$ and $k=1,\cdots,K-1$, we conclude the proof of Lemma 6.

## V. Conclusion

In this paper, we analyzed achievable DoF of the $M$-by-$N$ MIMO mRC with $L$ clusters and $K$ users per cluster, operating in the clustered full data exchange mode. We developed a novel systematic signal alignment technique to jointly construct the beamforming matrices at the users and the relay for efficient implementation of PNC. Based on that, an achievable DoF was derived for the considered MIMO mRC with an arbitrary configuration of $(L,K,M,N)$. We also showed that our proposed scheme is DoF-optimal for the considered MIMO mRC, when $\frac{M}{N} \leq \frac{1}{LK-1}$ and $\frac{M}{N} \geq \frac{(K-1)L+1}{KL}$.

The study of MIMO mRCs is still in an initial stage. The fundamental performance limits of such channels are far from being well understood. For example, the derived achievable DoF in this paper in general serves as a lower bound of the DoF capacity of the considered MIMO mRC. To narrow the gap towards the DoF capacity will be an important future research topic. Moreover, DoF analysis only characterizes the system performance at high SNR. Optimal beamforming designs for the MIMO mRC at finite SNR remains a challenging problem, and will be of interest for future research.

## Appendix A
## Some Useful Lemmas

Consider a full-rank matrix $\mathbf{A} = [\mathbf{A}_1, \cdots, \mathbf{A}_K]$, where $\mathbf{A}_i \in \mathbb{C}^{N\times M_i}$ for $i = 1, \cdots, K$. We make the following assumptions on the matrix size: $M_i \leq N, \forall i$, and $\sum_{i=1}^K M_i > N$. Then, $\mathbf{A}$ is a wide matrix with full row rank; also, each $\mathbf{A}_k$ is a tall matrix. Let $\mathbf{U} \in \mathbb{C}^{(\sum_{i=1}^K M_i)\times(\sum_{i=1}^K M_i - N)}$ be a nullspace matrix of $\mathbf{A}$, i.e., the columns of $\mathbf{U}$ give a basis of $\text{null}(\mathbf{A})$. Partition $\mathbf{U}$ as $\mathbf{U} = [\mathbf{U}_1^T, \cdots, \mathbf{U}_K^T]^T$ with $\mathbf{U}_i \in \mathbb{C}^{M_i \times (\sum_{i=1}^K M_i - N)}$. Denote $\widetilde{\mathbf{A}} = [\mathbf{A}_1\mathbf{U}_1, \cdots, \mathbf{A}_K\mathbf{U}_K] \in \mathbb{C}^{N\times K(\sum_{i=1}^K M_i - N)}$. Then

*Lemma 10:* Assume that $\mathbf{U}_k$ is of full rank, i.e., $\text{rank}(\mathbf{U}_k) = \min(M_i, \sum_{i=1}^K M_i - N)$, for $k = 1, \cdots, K$. Then, $\text{rank}(\widetilde{\mathbf{A}})$ is given by $\sum_{i=1}^K \min\left(M_i, \sum_{i=1}^K M_i - N\right) - \sum_{i=1}^K M_i + N$.

*Proof:* Let $\mathbf{v} = [\mathbf{v}_1^T, \cdots, \mathbf{v}_K^T]^T$ be an arbitrary vector in $\text{null}(\widetilde{\mathbf{A}})$, where $\mathbf{v}_i \in \mathbb{C}^{(\sum_{i=1}^K M_i - N)\times 1}$. Then, we obtain $\mathbf{0} = \widetilde{\mathbf{A}}\mathbf{v} = \mathbf{A}\text{diag}\{\mathbf{U}_1, \cdots, \mathbf{U}_K\}\mathbf{v}$. Thus, $\text{diag}\{\mathbf{U}_1, \cdots, \mathbf{U}_K\}\mathbf{v}$ belongs to $\text{null}(\mathbf{A})$. As $\mathbf{U}$ spans $\text{null}(\mathbf{A})$, there exists $\mathbf{x} \in \mathbb{C}^{(\sum_{i=1}^K M_i - N)\times 1}$ such that $\text{diag}\{\mathbf{U}_1, \cdots, \mathbf{U}_K\}\mathbf{v} = \mathbf{U}\mathbf{x}$, or equivalently, $\mathbf{U}_i\mathbf{v}_i = \mathbf{U}_i\mathbf{x}$, for $i = 1, \cdots, K$.

We now determine the number of free variables in $\mathbf{v}$ which is equal to the dimension of $\text{null}(\widetilde{\mathbf{A}})$. Note that $\mathbf{U}_i$ is of rank $\min(M_i, \sum_{i=1}^K M_i - N)$ for $i = 1, \cdots, K$. We consider the following two cases. First, if $M_i \geq \sum_{i=1}^K M_i - N$, the left inverse of $\mathbf{U}_i$ exists. Then, $\mathbf{U}_i\mathbf{v}_i = \mathbf{U}_i\mathbf{x}$ implies $\mathbf{v}_i = \mathbf{x}$, or equivalently, $\mathbf{v}_i$ is uniquely determined by $\mathbf{x}$. Second, if $M_i < \sum_{j=1}^K M_j - N$, from $\mathbf{U}_i\mathbf{v}_i = \mathbf{U}_i\mathbf{x}$, each $\mathbf{v}_i$ has $\sum_{j=1}^K M_j - N - M_i$ free variables for any fixed $\mathbf{x}$. Combining these two cases, we see that each $\mathbf{v}_i$ has $\max(\sum_{j=1}^K M_j - N - M_i, 0)$ free variables for any given $\mathbf{x}$. Considering all $\{\mathbf{v}_i\}$, together with the fact that $\mathbf{x} \in \mathbb{C}^{(\sum_{i=1}^K M_i - N)\times 1}$ is arbitrary, we obtain that $\mathbf{v}$ has

$$\sum_{i=1}^K M_i - N + \sum_{i=1}^K \max\left(\sum_{j=1}^K M_j - N - M_i, 0\right) \tag{49}$$

free variables, or equivalently, the dimension of $\text{null}(\widetilde{\mathbf{A}})$ is given by (49). By using the rank-nullity theorem, we conclude that the rank of $\widetilde{\mathbf{A}}$ is given by

$$\sum_{i=1}^K \min\left(M_i, \sum_{i=1}^K M_i - N\right) - \sum_{i=1}^K M_i + N, \tag{50}$$

which concludes the proof. ■

As an application of Lemma 10, we have the following result.

*Lemma 11:* Assume that $\mathbf{A}_i \in \mathbb{C}^{N\times M_i}$ for $i = 1, 2$ are random matrices with the coefficients independently drawn from an arbitrary continuous distribution. Also assume $M_i \leq N$ for $\forall i$ and $M_1 + M_2 > N$. The intersection of $\text{span}(\mathbf{A}_1)$ and $\text{span}(\mathbf{A}_2)$, denoted by $\text{span}(\mathbf{A}_1) \cap \text{span}(\mathbf{A}_2)$, is a subspace of dimension $(M_1 + M_2 - N)^+$ with probability one.

*Proof:* From the randomness of $\mathbf{A}_1$ and $\mathbf{A}_2$, $\text{null}([\mathbf{A}_1, \mathbf{A}_2])$ is of dimension $(M_1 + M_2 - N)^+$ with probability one. Let $\mathbf{U} = [\mathbf{U}_1^T, \mathbf{U}_2^T]^T \in \mathbb{C}^{(M_1+M_2)\times(M_1+M_2-N)^+}$ be a nullspace matrix of $[\mathbf{A}_1, \mathbf{A}_2]$. Again from the randomness of $\mathbf{A}_1$ and $\mathbf{A}_2$, $\mathbf{U}_1$ and $\mathbf{U}_2$ are of full rank for sure. From Lemma 10, we obtain that $\text{span}(\mathbf{A}_1\mathbf{U}_1, \mathbf{A}_2\mathbf{U}_2)$ is of dimension $\min((M_1 + M_2 - N)^+, N)$ for sure. Then, Lemma 11 follows by noting $\text{span}(\mathbf{A}_1) \cap \text{span}(\mathbf{A}_2) = \text{span}(\mathbf{A}_1\mathbf{U}_1, \mathbf{A}_2\mathbf{U}_2)$. ■

We next discuss the maximum number of independent units allowed for a pattern.

*Lemma 12:* For any pattern in (34), (39), (42), and (44), the maximum number of linearly independent units allowed is given by $\min(M, r)$, where $r$ is the nullspace dimension of the pattern matrix.

*Proof:* We focus on the pattern $\mathbf{B}_1\mathbf{u} = \mathbf{0}$ in (34). The proofs for the other patterns are similar and thus omitted for brevity. From the channel randomness, the rank of $\mathbf{B}_1$ is $(lKM, (l-1)N)$ for sure. From the rank-nullity theorem, $\text{null}(\mathbf{B}_1)$ is of dimension $r = (lKM - (l-1)N)^+$ for sure. Let $\mathbf{U} = [\mathbf{U}_{j_{[1]}}^T, \mathbf{U}_{j_{[2]}}^T, \cdots, \mathbf{U}_{j_{[l]}}]^T \in \mathbb{C}^{lKM\times r}$ be a nullspace matrix with the columns spanning the nullspace of $\mathbf{B}_1$. Further partition each $\mathbf{U}_{j_{[i]}} \in \mathbb{C}^{KM\times r}$ as $\mathbf{U}_{j_{[i]}} = [\mathbf{U}_{j_{[i]}1}^T, \cdots, \mathbf{U}_{j_{[i]}K}^T]^T$, where $\mathbf{U}_{j_{[i]}k} \in \mathbb{C}^{M\times r}$. Denote $\mathbf{M}_{j_{[i]}} = [\mathbf{H}_{j_{[i]}1}\mathbf{U}_{j_{[i]}1}, \cdots, \mathbf{H}_{j_{[i]}K}\mathbf{U}_{j_{[i]}K}]$, and $\widetilde{\mathbf{M}} = [\mathbf{M}_{j_{[1]}}, \cdots, \mathbf{M}_{j_{[l]}}]$. From the discussions below (34), each unit spans a subspace of dimension $lK - l + 1$. To construct $r$ linearly independent units as stated in Lemma 12, it suffices to show $\text{rank}(\widetilde{\mathbf{M}}) = (lK - l + 1)r$.

We first show that there exists such a nullspace matrix $\mathbf{U}$ that every component $\mathbf{U}_{jk}$ is of full rank, i.e., $\text{rank}(\mathbf{U}_{jk}) = \min(M, r)$ for $\forall j, k$. To see this, we rewrite (34) as

$$\mathbf{H}_{j_{[1]}}\mathbf{U}_{j_{[1]}} = -\mathbf{H}_{j_{[2]}}\mathbf{U}_{j_{[2]}} = \cdots = (-1)^{l-1}\mathbf{H}_{j_{[l]}}\mathbf{U}_{j_{[l]}} \tag{51}$$

or equivalently

$$\text{span}(\mathbf{H}_{j_{[i]}}\mathbf{U}_{j_{[i]}}) = \text{span}(\mathbf{H}_{j_{[1]}}) \cap \text{span}(\mathbf{H}_{j_{[2]}}) \cap \cdots \text{span}(\mathbf{H}_{j_{[l]}}) \tag{52}$$

for $i = 1, \cdots, l$. Recursively using Lemma 11, we see that the dimension of $\mathcal{S} = \text{span}(\mathbf{H}_{j_{[1]}}) \cap \text{span}(\mathbf{H}_{j_{[2]}}) \cap \cdots \text{span}(\mathbf{H}_{j_{[l]}})$ is of $\min(KM, r)$ for sure. As $\mathbf{H}_{j_{[1]}}, \cdots, \mathbf{H}_{j_{[l]}}$ are randomly generated, their intersection $\mathcal{S}$ is also random in $\mathbb{C}^N$. Further, for any index $i \in \{1, \cdots, l\}$, $\mathcal{S}$ is a random subspace of dimension $\min(KM, r)$ in $\text{span}(\mathbf{H}_{j_{[i]}})$. From linear algebra, $\mathbf{U}_{j_{[i]}} = [\mathbf{U}_{j_{[i]}1}^T, \cdots, \mathbf{U}_{j_{[i]}K}^T]^T$ is the coordinate matrix to describe a basis of $\mathcal{S}$ in $\text{span}(\mathbf{H}_{j_{[i]}})$. From the randomness of $\mathcal{S}$, $\mathbf{U}_{j_{[i]}k}$ for $k = 1, \cdots, K$ are of full rank for sure.

We are now ready to determine $\text{rank}(\widetilde{\mathbf{M}})$. Let $\mathbf{v} = [\mathbf{v}_{j_{[1]}}^T, \cdots, \mathbf{v}_{j_{[l]}}^T]^T$ be a vector in $\text{null}(\widetilde{\mathbf{M}})$, where $\mathbf{v}_j \in \mathbb{C}^{Kr\times 1}$.

To show $\text{rank}(\widetilde{\mathbf{M}}) = (lK - l + 1)r$, it suffices to show that $\text{null}(\widetilde{\mathbf{M}})$ is of dimension $(l-1)r$, or equivalently, $\mathbf{v}$ has $(l-1)r$ free variables. Let

$$\mathbf{v}'_{j_{[1]}} = \mathbf{v}_{j_{[1]}} \text{ and } \mathbf{v}'_{j_{[i]}} = \mathbf{v}_{j_{[i]}} - \mathbf{v}'_{j_{[i-1]}}, \text{ for } i = 2, \cdots, l. \quad (53)$$

From $\mathbf{B}_1\mathbf{U} = \mathbf{0}$, we obtain $\mathbf{H}_{j_{[i]}}\mathbf{U}_{j_{[i]}} + \mathbf{H}_{j_{[i+1]}}\mathbf{U}_{j_{[i+1]}} = \mathbf{0}$, for $i = 1, \cdots, l-1$. Thus

$$\mathbf{H}_{j_{[i]}}\mathbf{U}_{j_{[i]}}\mathbf{v}'_{j_{[i]}} + \mathbf{H}_{j_{[i+1]}}\mathbf{U}_{j_{[i+1]}}\mathbf{v}'_{j_{[i]}} = \mathbf{0} \quad (54)$$

for $i = 1, \cdots, l-1$, or equivalently

$$\widetilde{\mathbf{M}}\mathbf{v}''_{j_{[i]}} = \mathbf{0}, \text{ for } i = 1, \cdots, l-1 \quad (55)$$

where $\mathbf{v}''_{j_{[i]}} = [\mathbf{0}, \cdots, \mathbf{0}, \mathbf{v}'^T_{j_{[i]}}, \mathbf{v}'^T_{j_{[i]}}, \mathbf{0}, \cdots, \mathbf{0}]^T \in \mathbb{C}^{lKr\times 1}$ is a block-vector with only the $i$-th and $(i+1)$-th block-entries being nonzero. As $\mathbf{v} \in \text{null}(\widetilde{\mathbf{M}})$, we have $\widetilde{\mathbf{M}}\mathbf{v} = \mathbf{0}$; also, by definition, we have $\mathbf{v} = \sum_{i=1}^{l} \mathbf{v}''_{j_{[i]}}$. Subtracting the $l-1$ equations in (55) from $\widetilde{\mathbf{M}}\mathbf{v} = \mathbf{0}$, we obtain $\mathbf{H}_{j_{[l]}}\mathbf{U}_{j_{[l]}}\mathbf{v}'_{j_{[l]}} = \mathbf{0}$, which implies $\mathbf{v}'_{j_{[l]}} = \mathbf{0}$ by noting that both $\mathbf{H}_{j_{[l]}}$ and $\mathbf{U}_{j_{[l]}}$ are full column rank.[5] Therefore, only $\mathbf{v}'_{j_{[i]}}$ for $i = 1, \cdots, l-1$ may contain free variables.

Further denote $\mathbf{v}'_j = [\mathbf{v}'^T_{j1}, \cdots, \mathbf{v}'^T_{jK}]^T$, where $\mathbf{v}'_{j1} \in \mathbb{C}^{r\times 1}$. Then, rewrite (54) as

$$\sum_{k=1}^{K} \mathbf{H}_{j_{[i]}k}\mathbf{U}_{j_{[i]}k}\mathbf{v}'_{j_{[i]}k} + \sum_{k=1}^{K} \mathbf{H}_{j_{[i+1]}k}\mathbf{U}_{j_{[i+1]}k}\mathbf{v}'_{j_{[i]}k} = \mathbf{0} \quad (56)$$

for $i = 1, \cdots, l-1$. For each $i$, letting $\widetilde{\mathbf{A}} = [\mathbf{H}_{j_{[i]}1}\mathbf{U}_{j_{[i]}1}, \cdots \mathbf{H}_{j_{[i]}K}\mathbf{U}_{j_{[i]}K}, \mathbf{H}_{j_{[i+1]}1}\mathbf{U}_{j_{[i+1]}1}, \cdots, \mathbf{H}_{j_{[i+1]}K}\mathbf{U}_{j_{[i+1]}K}]$ and closely following the arguments in the proof of Lemma 10, we see that $\{\mathbf{v}'_{j_{[i]}k}\}$ must satisfy

$$\mathbf{U}_{j_{[i]}k}\mathbf{v}'_{j_{[i]}k} = \mathbf{U}_{j_{[i]}k}\mathbf{x}_i, \text{ for } k = 1, \cdots, K \quad (57)$$

where $\mathbf{x}_i$ is a vector in $\mathbb{C}^{r\times 1}$.

Consider the case of $r \leq M$, which implies that each $\mathbf{U}_{j_{[i]}k} \in \mathbb{C}^{M\times r}$ is of full column rank for sure. Then, as the right inverse of each $\mathbf{U}_{j_{[i]}k}$ exists, we obtain from (57) that $\mathbf{v}'_{j_{[i]}1} = \cdots = \mathbf{v}'_{j_{[i]}K}$. Thus, each $\mathbf{v}'_{j_{[i]}}$ contains $r$ free variables. Considering $\mathbf{v}'_{j_{[1]}}, \cdots, \mathbf{v}'_{j_{[l]}}$, we see that $\mathbf{v}$ have $(l-1)r$ free variables, i.e., the dimension of $\text{null}(\widetilde{\mathbf{M}})$ is given by $(l-1)r$. Therefore, $r$ units can be constructed.

Now consider $r > M$. From the above discussion, we are already able to construct $M$ units when $r = M$. Further increasing $r$ cannot support any more unit, for that in the considered mRC each user, equipped with $M$ antennas, is able to transmit at most $M$ independent signal streams. This concludes the proof. ■

## Acknowledgment

The author would like to thank Dr. Rui Wang for insightful discussions. The author would also like to thank the anonymous reviewers for their constructive suggestions to improve the presentation of this paper.

[5]Note that $r = lKM - (l-1)N \geq KM$ implies $\frac{M}{N} \geq \frac{1}{K}$, which exceeds the range of $\frac{M}{N}$ considered in constructing pattern (34). Thus, we have $lKM - (l-1)N < KM$, which implies $\mathbf{U}_{j_{[i]}} \in \mathbb{C}^{KM\times r}$ is a wide matrix. Then, from channel randomness, $\mathbf{U}_{j_{[i]}}$ is of full column rank.